%% file: main.tex
\documentclass[journal,twoside,web]{ieeecolor}
\usepackage{tikz,tikz-qtree}
\usepackage{generic}
\usepackage{cite}
\usepackage[hyphens]{url}
\usepackage[hidelinks]{hyperref}
\hypersetup{breaklinks=true}
\usepackage{amsmath,amssymb,amsfonts}
\usepackage{algorithmic}
\usepackage{graphicx}
\usepackage{multirow}
\usepackage{textcomp}
\usepackage{booktabs}
\usepackage{longtable}
\usepackage{makecell}
\usepackage{color, colortbl}
\definecolor{Gray}{gray}{0.9}
\definecolor{LightCyan}{rgb}{0.88,1,1}

\def\BibTeX{{\rm B\kern-.05em{\sc i\kern-.025em b}\kern-.08em
    T\kern-.1667em\lower.7ex\hbox{E}\kern-.125emX}}
\begin{document}
\title{Multi-modal Point-of-Care Diagnostics for {COVID-19} Based On Acoustics and Symptoms}
\author{Srikanth Raj Chetupalli,  \IEEEmembership{Member, IEEE}, Prashant Krishnan \IEEEmembership{Student Member, IEEE}, Neeraj Sharma, Ananya Muguli, Rohit Kumar, Viral Nanda, Lancelot Mark Pinto,  Prasanta Kumar Ghosh,  \IEEEmembership{Senior Member, IEEE} and  Sriram Ganapathy, \IEEEmembership{Senior Member, IEEE.}\thanks{Manuscript first submitted on June 1, 2021. This work was supported in part by the grants from the Department of Science and Technology, Government of India under the RAKSHAK program.}
\thanks{ Srikanth Raj Chetupalli, Prashant Krishnan, Neeraj Sharma, Ananya Muguli, Rohit Kumar, Prasanta Kumar Ghosh and Sriram Ganapathy are with Electrical Engineering, Indian Institute of Science, Bangalore 560012, INDIA.
Viral Nanda and Lancelot Mark Pinto are with the P. D. Hinduja National Hospital and Medical Research Center, Mumbai 400016, INDIA.}
\thanks{Corresponding author - Sriram Ganapathy (email:sriramg@iisc.ac.in).}}
\maketitle

\begin{abstract}
The research direction of identifying acoustic bio-markers of respiratory diseases has received renewed interest following the onset of COVID-19 pandemic. In this paper, we design an approach to COVID-19 diagnostic using crowd-sourced multi-modal data. The data resource, consisting of acoustic signals like cough, breathing, and speech signals, along with the data of symptoms, are recorded using a web-application over a period of ten months. We investigate the use of statistical descriptors of simple time-frequency features for acoustic signals and binary features for the presence of symptoms. Unlike previous works, we primarily focus on the application of simple linear classifiers like logistic regression and support vector machines for acoustic data while decision tree models are employed on the symptoms data.
We show that a multi-modal integration of acoustics and symptoms classifiers achieves an area-under-curve (AUC) of $92.40$, a significant improvement over any individual modality.
Several ablation experiments are also provided which highlight the acoustic and symptom dimensions that are important for the task of COVID-19 diagnostics.
\end{abstract}


\section{Introduction}\label{sec:introduction}
\IEEEPARstart{A}{} highly contagious variant of the coronavirus family, SARS-CoV-2 has resulted in the most  significant  health crisis of the  twenty-first century \cite{hu2020characteristics}. The outbreak was termed as the coronavirus disease 2019 (or COVID-19) and declared a pandemic in March 2020 by the World Health Organization (WHO). While vaccination efforts have partly reduced the viral spread in some parts of the world, the multiple waves of infections across various countries indicate that the COVID-19 pandemic has the potential to persist for months and years to come \cite{hu2020characteristics}.

The pathogenesis of COVID-19 suggests that infection triggers a series of events that enable the SARS-CoV-2 virus to replicate and migrate down the respiratory tract to the epithelial cells in the lungs~\cite{hu2020characteristics}. 
The most common timeline for these events is $2$-$5$ days from the onset of the infection. 
Further, the risk of viral spread to primary contacts is the highest during the first week of infection. Early diagnosis can help to identify, isolate infected individuals, and control the community spread of the virus~\cite{who_test_test}.  

\subsection{Current tests and limitations}
\noindent Currently, the gold-standard of COVID-19 diagnosis is the reverse transcription polymerase chain reaction (RT-PCR) assay~\cite{corman2020detection}. The RT-PCR, based on molecular testing, detects the amino-acid sequences unique to SARS-CoV-2 in swab samples. The throat and/or nasal swab samples are first collected, stored, and processed at a lab facility, where the ribo-nucleic acid (RNA) content is amplified for detecting the presence of the COVID-19 genome in the sample. However, the RT-PCR has four major limitations when it comes to massive population level scaling: i) the cost of RT-PCR chemical reagent and facility, ii) expert supervision, iii) the turnaround time from sample collection to results (hours to days), and iv) lack of physical distancing during sample collection.

A widely used alternative to RT-PCR testing is the rapid antigen testing (RAT) methodology~\cite{peeling2021scaling}. In particular, the RAT attempts to identify one of the outer proteins of the viral shell or envelope with results available in $15-20$~min. This testing methodology is less expensive compared to the chemical reagent based testing. However, the key limitation of potential spread during sample collection is not alleviated. Further, the sensitivity values for the same level of specificity are lower compared to RT-PCR tests~\cite{scohy2020low}. 

In summary, there is a need to discover alternative test methodologies which improve the trade-off between time, cost, physical distancing, and performance.
The WHO blueprint on COVID-19 diagnostic tests highlights the urgent need for developing point-of-care tests (POCTs) \cite{who_poct}. 
In this paper, we present a study analyzing the respiratory acoustics of COVID-19 infection and its suitability  for the design of POCT tools.

\subsection{Acoustics for respiratory diagnostics}
\noindent Listening to acoustic changes in lung sounds for preliminary screening of abnormalities in the respiratory passage was first formalized by Laennec et al. \cite{laennec1838treatise} in $1838$. With the advances in sensor and communication technologies, ``machine listening'' approach for automatic analysis of respiratory sounds has gained interest. The evaluation metrics typically employed are area under the curve (AUC) and specificity at a pre-defined sensitivity. Pramono et. al.~\cite{pramono2016cough} developed an  automated diagnosis of pertussis using cough sound signals. A smartphone based multi-modal user-friendly approach to the detection of chronic obstructive pulmonary disease (COPD) and congestive heart failure was explored by Windmon et. al.~\cite{windmon2018tussiswatch}. Botha~et.al.~\cite{botha2018detection} showed that a spectral analysis of cough sounds can provide a low cost detection of tuberculosis. In a recent work, Porter et. al.~\cite{porter2020diagnosing} developed a smartphone based portable cough sound and reported symptoms analysis to detect chronic airway disease.
Additionally, in controlled studies using small cohort sizes, effectiveness of machine learning based sound analysis has been demonstrated for childhood pneumonia detection \cite{abeyratne2013cough}, wet versus dry cough classification \cite{swarnkar2013automatic}, and asthmatic versus healthy classification \cite{hee2019development}.
These efforts illustrate that acoustic signals have the potential to allow low-cost, rapid and efficient diagnostics of respiratory ailments. 

For COVID-19, a meta-analysis study of symptoms by Li et al.\cite{li2021epidemiology} found fever ($78.8\%$), followed by cough ($53.9\%$), and malaise ($37.9\%$) as the common symptoms in $281,641$ COVID-19 infected individuals. The  clinical  symptoms of COVID-19 include fever, common cold, cough, chest congestion, breathing difficulties, dyspnea, loss of smell/taste and pneumonia~\cite{HUANG2020497}. 
The human sounds, such as cough, breathing, and speech, result from a coordinated functioning of the lungs and the organs in the respiratory pathway. An impairment in the functioning of these organs, such as constriction, fatigue, or difficulty in breathing, may be manifested in the acoustic characteristics of these sounds. These observations motivate the design and evaluation of a sound sample based diagnosis approach for COVID-19~\cite{9391791}. A success can lead to creation and large-scale deployment of a POCT tool, where an individual can record their sound samples on  a portable web-connected device, and an application can analyze and display the result.

\begin{figure*}[t]
    \centering
    \includegraphics[width=0.9\linewidth, height=3.6in] {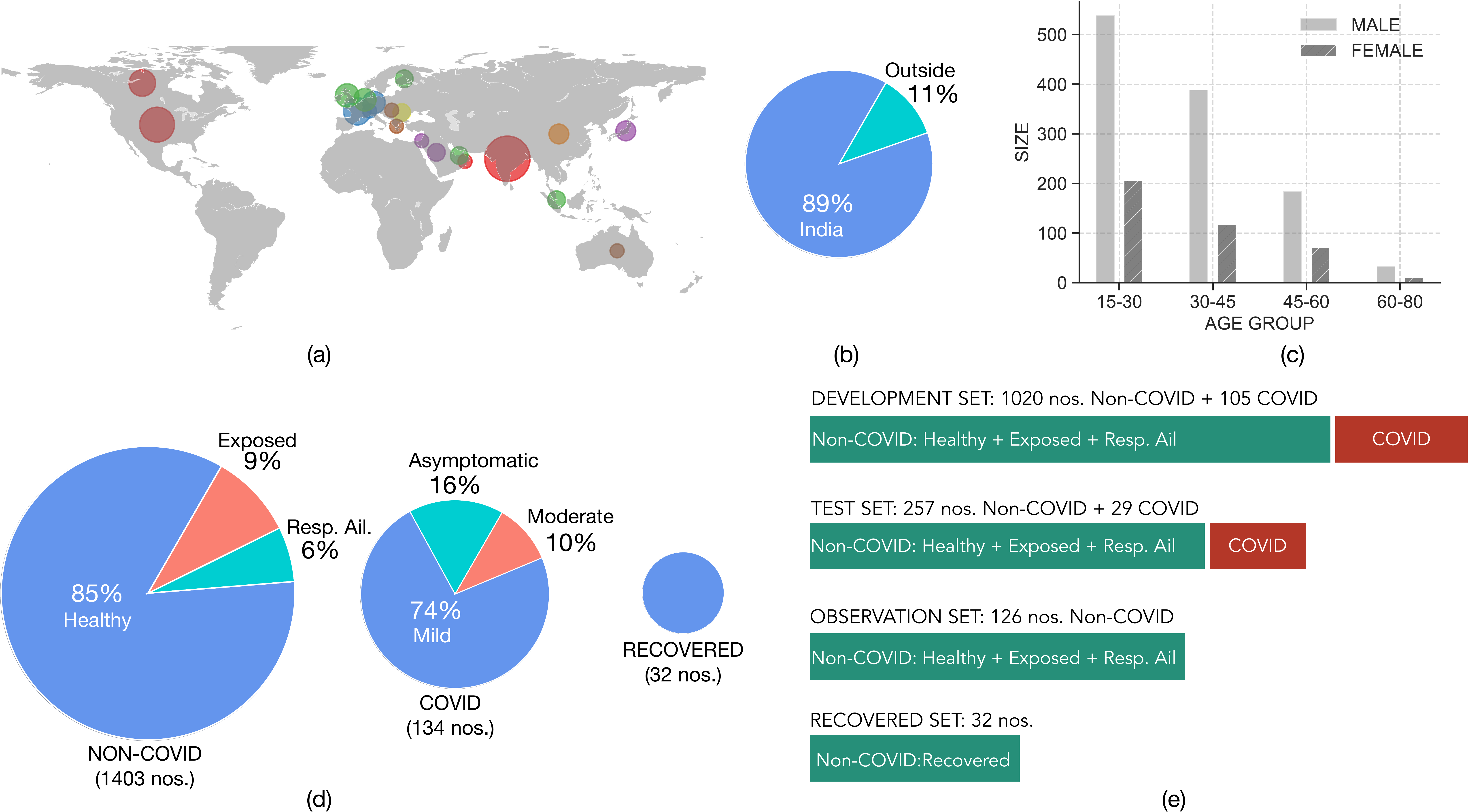}
    \caption{(a) Broad geographic distribution of participants, (b) Percentage of participants from India and outside, (c) age group and gender breakup, (d) grouping of participants into three pools: non-COVID, COVID, and Recovered, and (e) division of the three pools into development, test, observation, and recovered set for the proposed study.}
    \label{fig:age_gen_dist}
\end{figure*} 

\section{Related prior work and contributions}
\label{sec:lit_review}
\noindent   For sound sample dataset creation, the notable efforts include the COVID-19 Sounds project (cough and breathing) by Univ. Cambridge  \cite{brown2020exploring}, COUGHVID dataset (cough) by EPFL \cite{orlandic2020coughvid}, and the  COVID-19 Cough dataset by MIT \cite{laguarta}. 
Further, our team has also been involved in the dataset creation task~\cite{sharma2020coswara}. 


Using cough and breathing sounds, Brown et al. \cite{brown2020exploring} report a performance measure of $0.80$ AUC on a subset of COVID-19 Sounds dataset. On a similar dataset, Coppock et al. \cite{coppock2021end} report $0.85$ AUC.
Agbley et al. \cite{agbley2020wavelet} demonstrated a specificity of $0.81$ at a sensitivity of $0.43$ on a subset of COUGHVID dataset. Feng et al. \cite{9401826} used a subset of cough sounds from Coswara dataset and reported a performance of  $0.90$ AUC. Laguarte et al. \cite{laguarta} obtained AUC greater than $0.90$ on samples from the COVID-19 Cough data set. These studies use acoustic feature representations of cough sounds such as Mel frequency cepstral coefficients (MFCCs) \cite{brown2020exploring}, Mel-spectrogram \cite{coppock2021end,laguarta}, or scalograms \cite{agbley2020wavelet}, while the classifier models are deep learning based neural networks such as convolutional neural networks (CNNs) \cite{agbley2020wavelet}, recurrent neural networks (RNNs) \cite{9401826}, CNN based feature embeddings in support vector machines (SVM)  \cite{brown2020exploring} or with CNN based residual networks \cite{coppock2021end, laguarta}. There are also attempts at creating more controlled COVID-19 cough sound dataset from individuals in hospitals \cite{9361107, imran2020ai4}. 

For voice sounds, Verde et al. \cite{9416469} used sustained phonation of vowels $/a/, /e/, /o/$  from a subset of the Coswara dataset, and obtain AUCs in the range $0.71-0.97$ using different kinds of classifier.

Using self-reported symptoms of COVID-19, Menni et. al.~\cite{smell_metadata} showed that an  AUC of $0.74$ is achievable from two different sets of recorded data (from the US and UK). A recent study by Zaobi et. al.~\cite{zoabi2021machine} further extended the analysis using a large pool of COVID-19 and healthy subjects.
\subsection{Contributions}
\noindent The key contributions from the current work are as follows.
\begin{enumerate}
        \item Investigating a multi-modal integration approach to classification using cough, breathing and speech signals, along with self-reported symptoms. 
    \item Exploring recording level statistical feature descriptors of acoustic signals, and binary feature encoding of symptom data.
    \item Emphasis on simple linear classifier models like linear regression and support vector machine approaches for COVID-19 classification task.
    \item Understanding the importance of feature dimensions using ablation studies.
\end{enumerate}
\section{Materials}
\subsection{Dataset}\label{sec:dataset}
\noindent This study is based on sound and symptom samples  from the Coswara dataset\footnote{\url{https://github.com/iiscleap/Coswara-Data}}~\cite{sharma2020coswara}. We use the data recorded (and released) up to 07-May-2021. This is composed of contributions from $1699$ participants ($157$ COVID-19 positive).
Each participant contributes $9$ audio recordings, namely, $(a)$ shallow and deep breathing, $(b)$ shallow and heavy cough, $(c)$ sustained phonation of three vowels /ae/ (as in bat), /i/ (as in beet), and /u/ (as in boot), and $(d)$ fast and normal pace $1-20$ number counting. 
Alongside this, each participant also records current health status (COVID-19 infection, symptoms and co-morbidity, if any), gender, age, and broad geographical location. No personally  identifiable information is collected.
The dataset collection protocol is approved by the Human Ethics Committee of the Indian Institute of Science,  Bangalore and the P. D. Hinduja National Hospital and Medical Research Center, Mumbai, India. In this study, we focus on modeling and analysis of three sound categories, namely, (i) breathing-deep (breathing), (ii) cough-heavy (cough), and (iii) counting-normal (speech) and the symptom data.
 
In this study, we select participants in the age group of $15-80$ yrs. Any participant with less than $100$~ms of sound sample or peak amplitude less than $10^{-4}$ are removed. The resulting subset consists of data from $1569$ participants. An illustration of geographic distribution, age, and gender is shown in Figure~\ref{fig:age_gen_dist} (a,b,c). The participants come from several countries; however, $89\%$ belong to India. The majority of the participants belong to the $15-45$ years age group, and a majority are male ($73\%$). The $1569$ participants can be further grouped into three pools (Figure~\ref{fig:age_gen_dist}(d)).
In the first pool, referred to as non-COVID, $1403$ participants are self-declared COVID-19 negative. This comprises individuals who are healthy ($85\%$), exposed to COVID-19 positive patients $(9\%)$, or have pre-existing respiratory ailments.
In the second pool, referred to as COVID, $135$ participants are self-declared COVID-19 positive. This comprises individuals who have mild ($73\%$), moderate ($16\%$), or asymptomatic COVID-19 infection at the time of recording their audio sample. In the third pool, referred to as recovered, we have $32$ participants. 
\begin{figure}[t!]
    \centering
   \includegraphics[width=3.35in,height=1.48in]{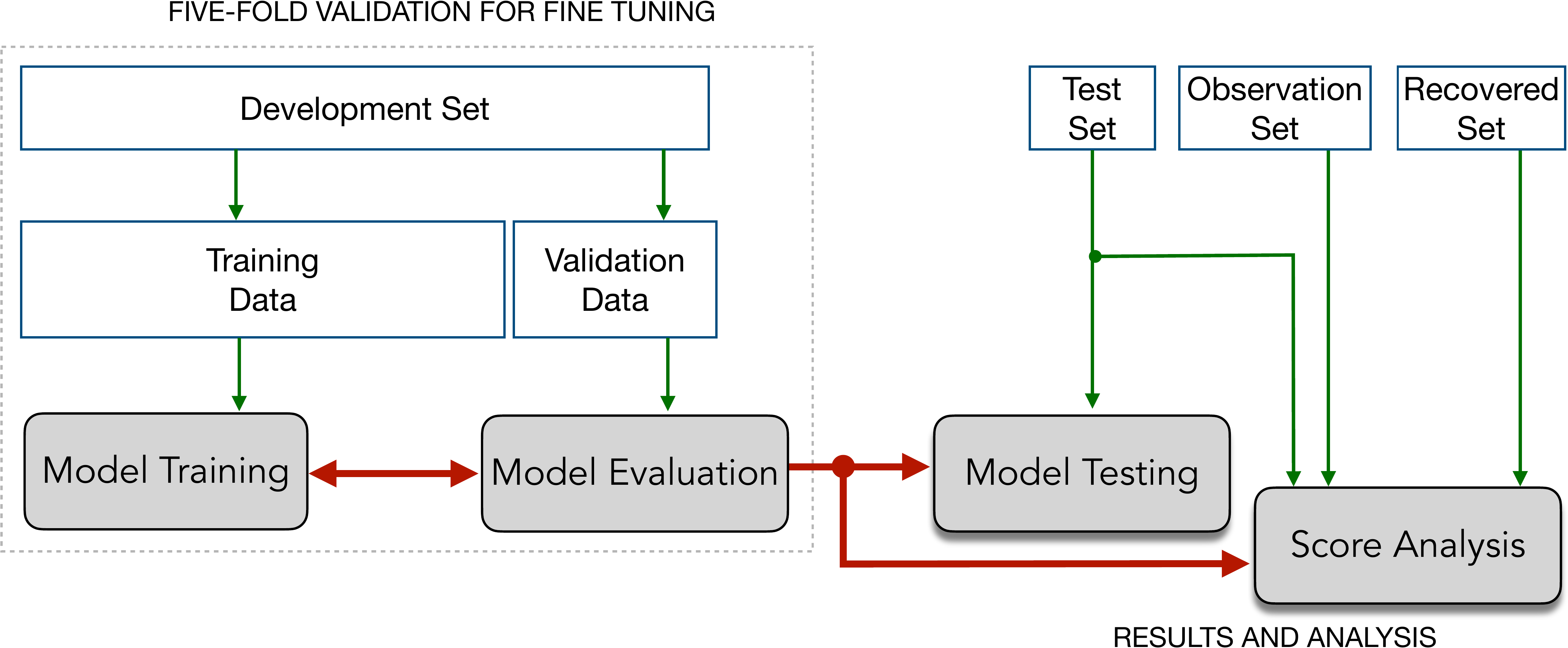}
    \vspace{-0.05in}
    \caption{The dataset modeling and analysis flow. The development data is split into training and validation data, and used for five-fold validation experiments. The test set is used for evaluating the performance metrics. The observation set and recovered sets are used for score analysis.}
    \vspace{-0.2in}
    \label{fig:bd_experiment_procedure}
\end{figure}
\subsection{Dataset division}
\noindent The three pools of participants, shown in Figure~\ref{fig:age_gen_dist}(d), are re-organized to create non-overlapping subsets facilitating development of classifiers, testing, and analysis (Figure~\ref{fig:bd_experiment_procedure}). The subsets are also illustrated in Figure~\ref{fig:age_gen_dist}(e).  
\subsubsection{Development and test set}
These are obtained by performing a $80-20\%$ random split on both non-COVID and COVID pools after  removing the observation set ($126$ participants). The resulting subsets are referred to as the development and test sets. The development set is composed of $1125$ ($106$ COVID) participants, and the test set has $286$ ($29$ COVID) participants. The development set is further divided into training and validation sets in a five-fold validation setup. The train/validation folds are used for training and fine-tuning the classifiers. All the subsets have  COVID and non-COVID pool distributions for gender/age groups similar to the full data.

\subsubsection{Observation set}
From the non-COVID pool shown in  Figure~\ref{fig:age_gen_dist}(d), we partition the  data recorded between April $1$-May $7$, $2021$. This data was recorded during the second wave of COVID-19 infections in India. This subset of non-COVID pool ($126$ nos.) is only used for score analysis in Section~\ref{sec:score_analysis}.
\subsubsection{Recovered set}
The set of participants who self-reported as recovered ($32$ participants) is also separated out of development and test sets. As the date of recovery is not collected during recording, we only  analyze the score distribution of these participants  in Section~\ref{sec:score_analysis}.

\section{Methods}\label{sec:methods}
\subsection{Classification models}
\begin{figure}[t!]
    \centering
    \includegraphics[width=2.95in,height=3.15in]{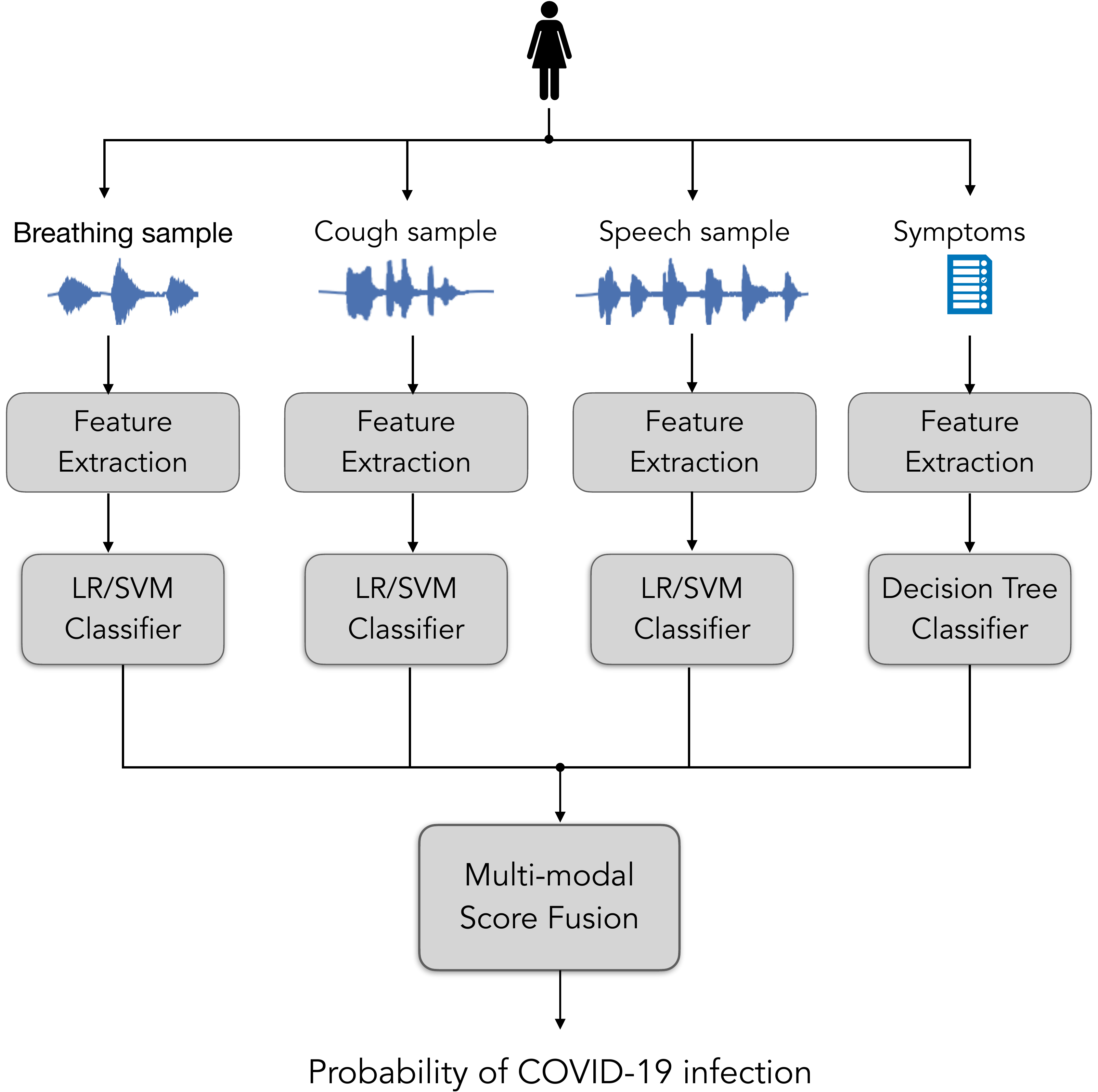}
    \vspace{-0.05in}
    \caption{Block schematic of the proposed multi-modal diagnostics for COVID-19 (MuDiCov) system.}
    \label{fig:block_MuDiCov}
\end{figure}

\noindent The block schematic of the multi-modal diagnostic tool proposed in this work is shown in Figure~\ref{fig:block_MuDiCov}. We consider the  binary classification task designed to separate COVID participants from non-COVID participants.
On the sound sample data, we explore the  logistic regression (LR) model, support vector machine with linear kernel (Lin-SVM) model and support vector machine with radial basis function kernel (RBF-SVM) model. On symptom data, a decision tree model is used. 

For the discussion below, let $\boldsymbol x$ denote an acoustic feature vector and $\boldsymbol y$ denote a symptom feature vector. We denote the prediction score (higher values indicating a higher probability of COVID class), an output of the model, by $p$. 

\textbf{Logistic regression (LR)}: The LR model generates the prediction score ($p$) as,
\begin{equation}
p = \sigma ({\boldsymbol w}^T {\boldsymbol x} + b)     
\end{equation}
where, ${\boldsymbol w}$ and $b$ are the weight vector and the bias of the model, respectively. Here, $\sigma$ is the logistic function, $\sigma(a) = (1+e^{-a})^{-1}$. The LR model is trained by minimizing the cost function $E(.)$ defined as, 
\begin{equation}\label{eq:lossLR} 
 E({\boldsymbol w},b) =  -[c~\log (p) + (1-c)~ \log(1-p)]  + \lambda ||\boldsymbol {w}||_{2}^2, 
\end{equation}
where, $c$ denotes the class label of the feature vector $x$, $c=1$ for COVID class and $c=0$ for non-COVID class, $||.||_{2}$ is the $\ell_2$-norm of the vector and $\lambda$ is a regularization parameter. The cost function is optimized using standard gradient based methods \cite{steepestdescent}.

\textbf{Support vector machine (SVM)}: The linear SVM (Lin-SVM) model generates prediction scores as,
\begin{equation}
 p = f\left({\boldsymbol w}^T {\boldsymbol x} + b\right)     
\end{equation}
where, ${\boldsymbol w}$ and $b$ are the weight vector and bias of the model, respectively; and $f()$ denotes the Platt scaling based calibration \cite{Platt99probabilisticoutputs}. The model is learned by minimizing the soft-margin cost function $E(.)$ defined as, 
\begin{equation}\label{eq:lossSVM}
 E({\boldsymbol w},b) = \max (0, 1-c ({\boldsymbol w}^T {\boldsymbol x} + b)) + \lambda ||\boldsymbol w||^2,     
\end{equation}
where, $c$ denotes the class label of the feature vector $x$, $c=1$ for COVID class and $c=-1$ for non-COVID class. The above cost function is optimized using constrained optimization methods involving primal-dual modeling. In the dual space, the weight vector $\boldsymbol w$ is expressed as a function of the inner-product of the feature matrix~\cite{Platt99probabilisticoutputs}. By replacing the inner-product with a kernel, the Lin-SVM model can be made to operate in a higher dimensional space. We explore SVM with radial basis function kernel (RBF) defined as,
\begin{equation}
    k({\boldsymbol x}_i, {\boldsymbol x}_j) = \exp\left(-\frac {|| {\boldsymbol x}_i - {\boldsymbol x}_j||^2}{\gamma}\right)
\end{equation}
where $k$ is the kernel function of two data points ${\boldsymbol x}_i$, ${\boldsymbol x}_j$ and $\gamma$ is a  free parameter of the RBF kernel. 

The acoustic features that are input to the classifier are standardized using the global mean-variance, computed on the training data. 

\textbf{Decision tree}:
Each  node in the tree is associated with a feature dimension. The edges drawn out of a node indicate the value of the feature dimension for each possible value (in our case, the features are binary). The leaf nodes are associated with a posterior probability distribution over the classes. In a classification tree model, the leaf nodes represent classification decisions while the other nodes represent the set of conditions applied on to the features that lead to the class labels. The Gini criterion is used by the classifier~\cite{myles2004introduction}, and the minimum number of samples at the leaf node in the tree is chosen based on the cross-validation.

\subsection{Performance metrics}
\noindent The primary metric used in our analysis is the area-under-the curve (AUC) measure of the receiver operating characteristic curve (ROC). The ROC plots ``1-specificity'' versus the sensitivity. The sensitivity (a.k.a true positive rate) and specificity (a.k.a false negative rate) are defined as,
\begin{eqnarray}
    \text{Sensitivity} &=& \frac{\mbox{\# correctly predicted COVID labels}}{\mbox{\# COVID labels}} ~~~~\\
    \text{Specificity} &=& \frac{\mbox{\# correctly predicted non-COVID labels}}{\mbox{\# non-COVID labels}}~~~
\end{eqnarray}
where, label stands for a participant. We compute the ROC curve by varying the decision threshold from $0$ to $1$ in steps of $10^{-4}$ and obtaining the specificity and sensitivity at each of these thresholds. The AUC is computed using the trapezoidal rule.  The positive predictive value (PPV) is the probability that a participant with a positive decision from the test has the COVID-19 infection. Similarly, negative predictive value (NPV) is the probability that a participant with a negative decision does not have the COVID-19 infection.

\subsection{Audio pre-processing}
\noindent The Coswara dataset provides sound samples as WAV format audio files. A majority ($>90\%$) of these are sampled at $44.1~$kHz. We standardize all sound files to a sampling rate of $44.1~$kHz via re-sampling. The amplitude range of the audio file is also normalized to $\pm 1$. Any initial and trailing silences in the audio files (greater than $50~$msec on either side) is removed using threshold amplitude value of $10^{-4}$. The average duration (and standard deviation) of sound samples correspond to $16.4~(\text{std.~dev.}~6.25)$~secs, $5~(\text{std.~dev.}~2.31)$~secs, and $13.6~(\text{std.~dev.}~4.1)$~secs for breathing, cough, and speech categories, respectively.

\subsection{Audio feature extraction}
\noindent  We make use of the low-level descriptors (LLDs) referred to as ComParE2016~\cite{ComParE2016} for this.
These descriptors broadly quantify the energy, spectral, and voicing attributes in an acoustic signal.
These are listed in Table~\ref{tab:llds}. We further quantify statistical properties of each of the LLDs over time (and/or frequency).
There are approximately $100$ statistical measures quantified for each LLD.
A broad categorization of these properties is listed in Table~\ref{tab:functionals}. 
A detailed description of the LLDs and the estimated statistical measures is available in the reading  material\footnote{\url{https://github.com/iiscleap/MuDiCov/blob/main/ComParE2016_Featureset_Description.pdf}}.  Every sound sample is represented by a $6373$ dimensional   feature vector. 
The same set of features are derived for all  categories of the acoustic signals namely, cough, breathing and speech. 
\input{plots/tables/LLDs}
\input{plots/tables/Functionals}

From Table \ref{tab:llds}, and \ref{tab:functionals}, we see that the feature set is a rich set of redundant features, with the multiple descriptors capturing similar acoustic properties of the audio signal. However, since the features are well defined functions on low-level acoustic descriptors, the analysis using simple linear models of classification allows us to understand the acoustic characteristics of the cough, speech and breathing signals that enable the classification of COVID and non-COVID  participants. This analysis is given in Section~\ref{sec:score_analysis}.

\subsection{Features from symptoms}
\begin{figure}[t!]
    \centering
    \includegraphics[width=3.15in, height=2.36in]{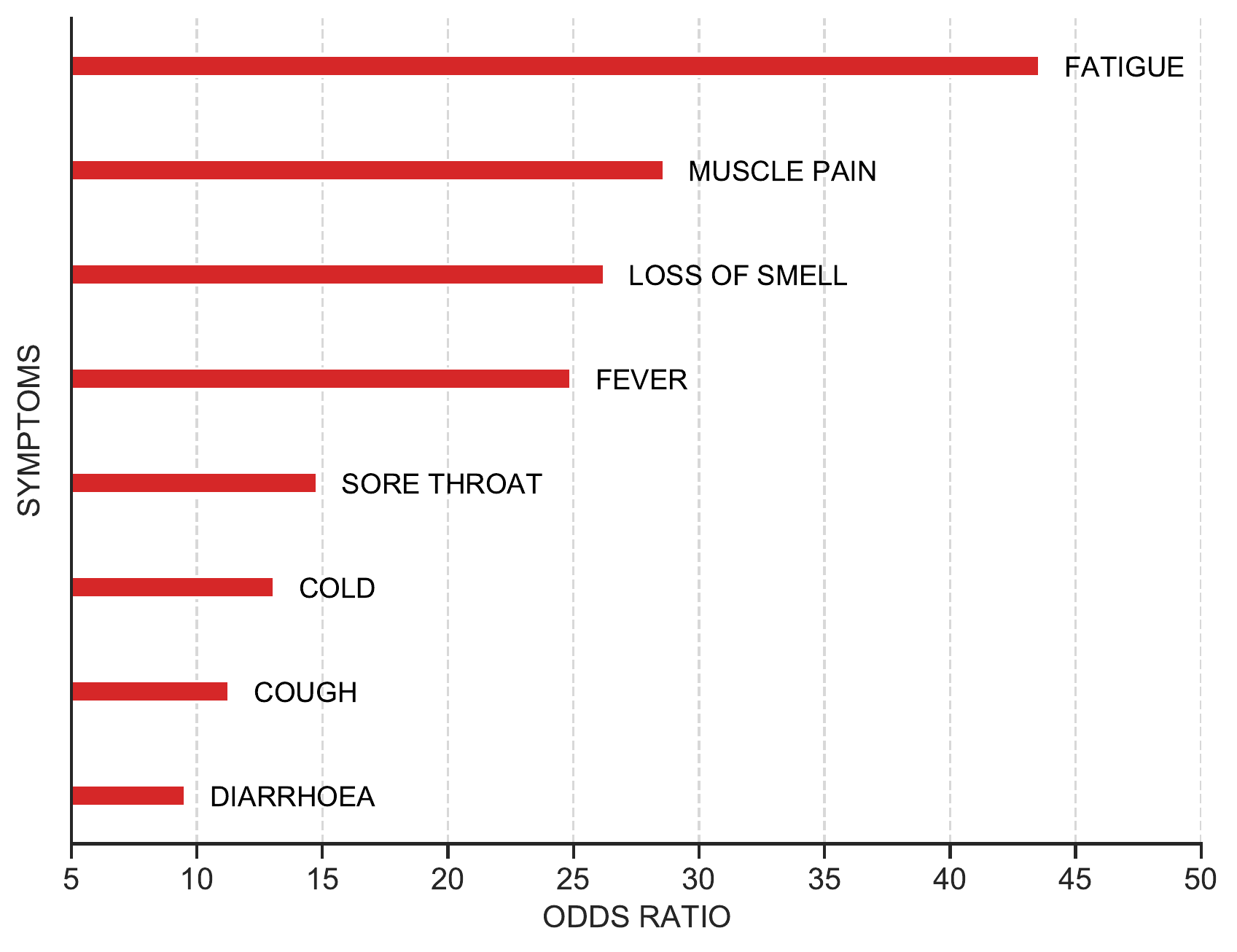}
    \caption{Odds ratio of the symptoms data in pooled development and test set.}
    \label{fig:symptoms_distribution}
    \vspace{-0.15in} 
\end{figure}
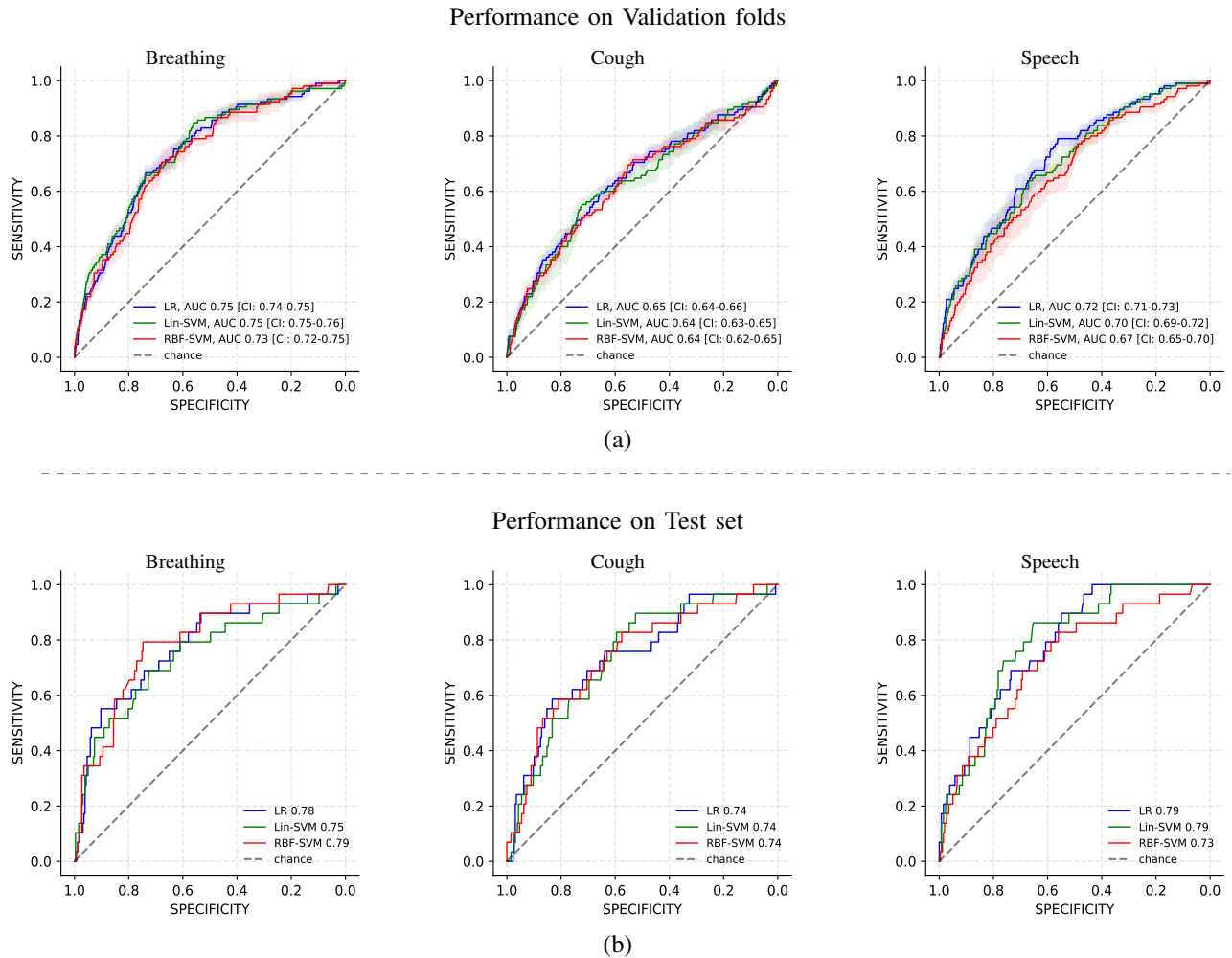
\begin{figure*}[t!]
\centering
\input{val_test_plots}
    \caption{(a) Top row shows the ROC curves obtained for the five fold cross validation on the development data. (b) Bottom row shows the ROC curves for test set using the model trained on the entire development set. }
    \label{fig:functionals_summary}
\end{figure*}
\noindent The dataset also has information on the presence/absence of $8$ common symptoms from all the participants. The odds ratio of the participants with  symptoms among the COVID and non-COVID category is shown in Figure \ref{fig:symptoms_distribution}. Not all COVID participants have symptoms, and a few participants have more than one symptom. In terms of proportions, a higher proportion of participants with COVID have symptoms than the non-COVID participants. Figure \ref{fig:symptoms_distribution} shows that the odds-ratio is higher for fatigue, muscle pain and loss of smell.

In our model development, the symptoms are converted to $8$ dimensional binary features for each participant. Each feature dimension  is set to $1$ or $0$ depending on the presence or absence of the symptom respectively.  The binary vectors derived in this manner are used to train and test a classifier which is input with these features. 

\subsection{Multi-modal fusion}\label{sec:scorefusion}
\noindent We explore the fusion of predicted probability scores from multiple modalities of acoustic data (cough, breathing and speech) and symptom data. Let $\{p_1,p_2,...,p_N\}$ be the scores obtained from $N$ different classifiers (maximum value of $N$ is $4$) for a given participant. The fused score is,
\begin{equation}\label{eq:fusion}
    p = \frac{1}{N} \sum_{i=1}^{N} p_i.
\end{equation}

\subsection{Implementation}
\noindent We use the OpenSmile~\cite{opensmile} Python toolbox\footnote{\url{https://github.com/audeering/opensmile-python}} to extract the features from sound samples.
The classifiers are implemented using the Scikit-learn Python toolkit \cite{scikit-learn}.   In order to allow reproducible research, all the implementation scripts to extract features and train the classifiers reported in this work are available at \url{https://github.com/iiscleap/MuDiCov}.
\input{plots/decision_tree}

\begin{figure*}[t!] 
    \centering
    \includegraphics[width=7.25in,height=2.75in]{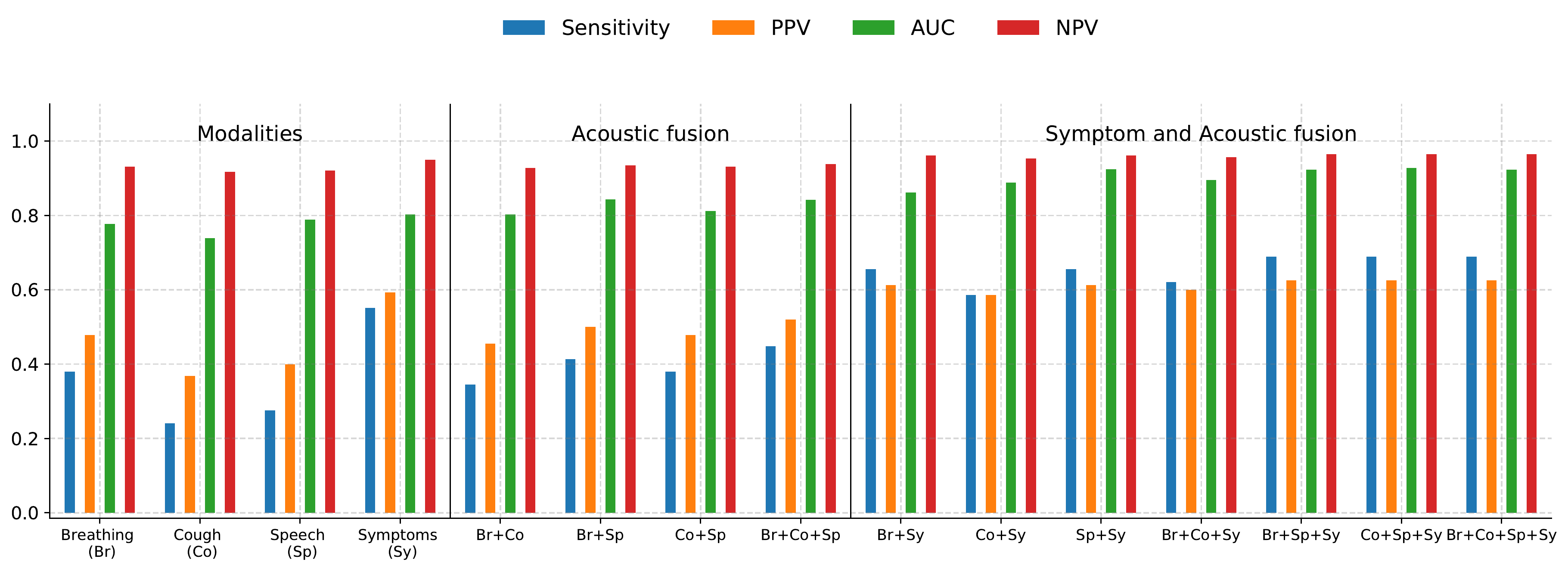}
    \caption{Performance measures for the individual modalities and score fusion of all combinations of modalities. Here the sensitivity, PPV and NPV values are measured for a specificity of $95$\%}
    \label{fig:functionals_score_combination}
\end{figure*}
\begin{figure}[t!]
    \centering
    \includegraphics[width=3.34in,height=3.00in]{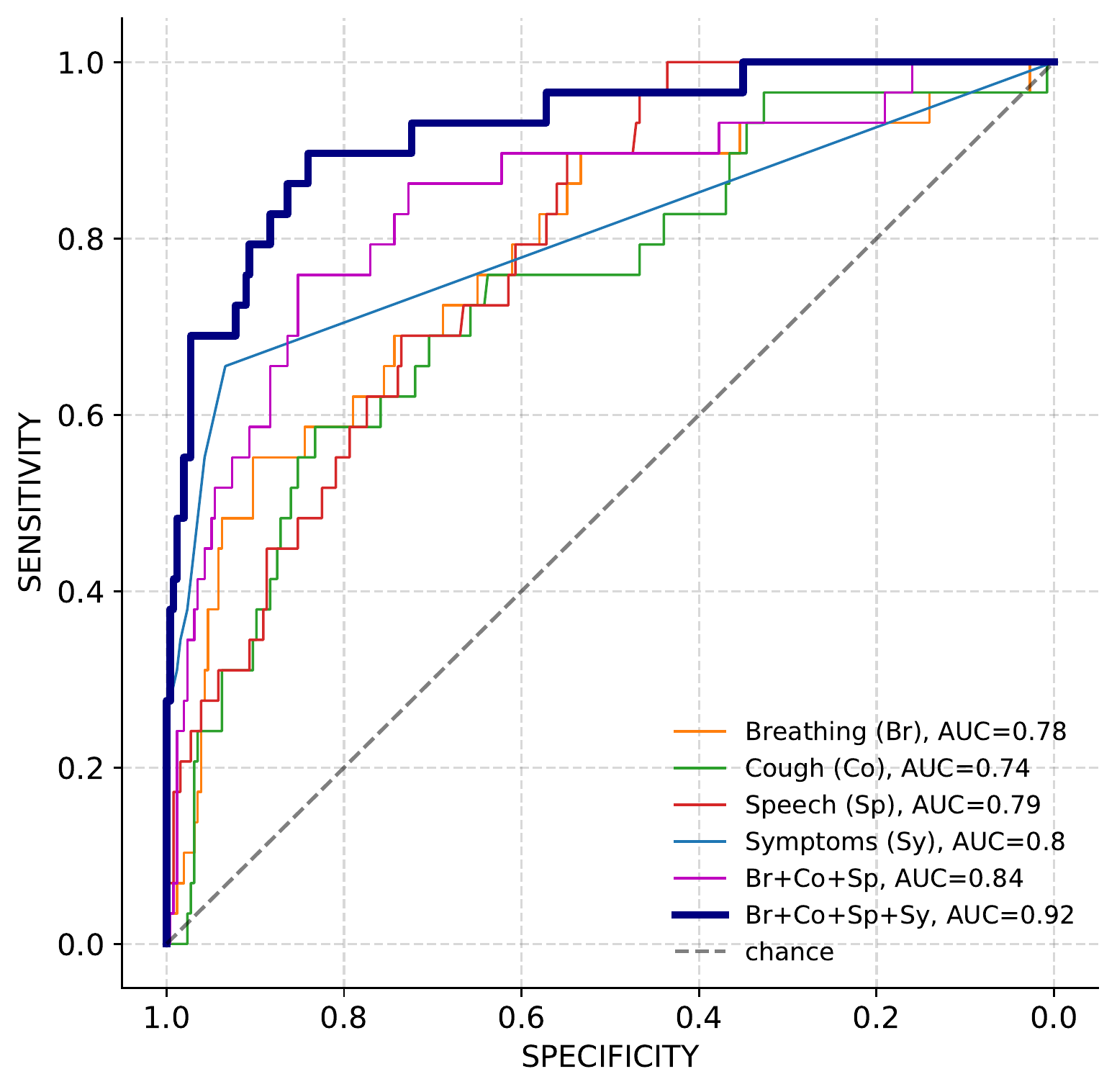}
    \vspace{-5pt}
    \caption{Test ROCs of the individual and the fusion systems for the LR classifier.}
    \label{fig:scorefusion_rocs}
    \vspace{-0.2in}
\end{figure}
\section{Experiments and Results}
\noindent The experiments are performed on the dataset splits shown in Figure \ref{fig:bd_experiment_procedure}. The hyper-parameters, namely, $\lambda$ in Equation~\eqref{eq:lossLR} (LR) and Equation~\eqref{eq:lossSVM} (SVM), and minimum number of samples in leaf nodes (decision tree), are selected using a five-fold validation procedure on the development data. The final value for each hyper-parameter is chosen based on the best average AUC measure on the five-folds. Subsequently, the classifier is   trained on the entire development set with the chosen hyper-parameter value. This classifier is evaluated on the test set. We use the default options for other configuration parameters. For all the classifiers, we use the ``balanced loss'' option in the classifier configuration. In the balanced loss setting, the loss value for the positive class samples is weighted by the ratio of the number of negative class samples to positive class samples in the training set. 

\subsection{Acoustic classifiers}
\noindent The top row in Figure \ref{fig:functionals_summary} depicts the five-fold cross validation results obtained for the three classifiers, independently trained on each of the three acoustic modalities. The plots correspond to the classifier with the best regularization parameter $\lambda$, identified based on maximum average validation AUC. The breathing (Br) and speech (Sp) categories performed relatively better compared to the cough (Co) data. The confidence interval around the mean ROCs and the mean AUCs indicates that the variance across the folds is also small.  

\begin{figure*}[t!]
    \centering
    \includegraphics[width=7in,height=1.53in]{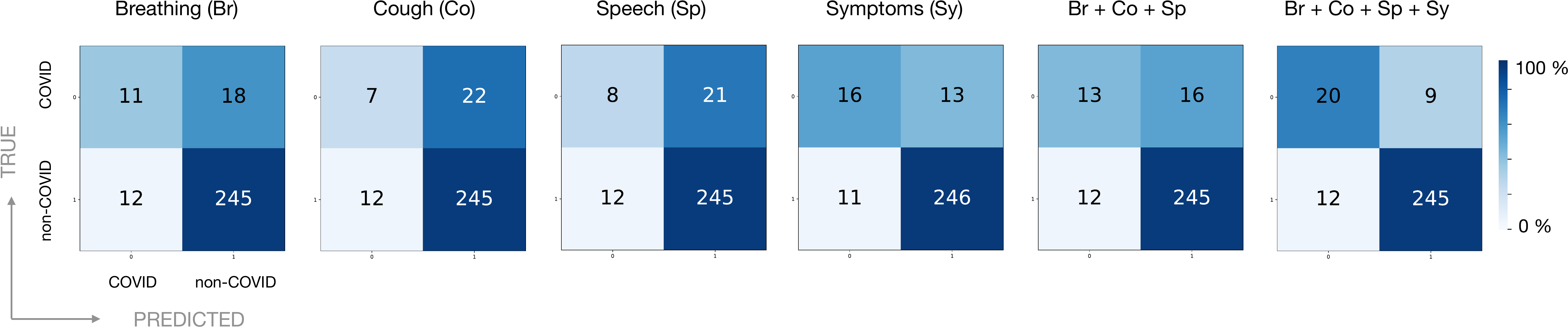}
    \caption{Confusion matrices on test data, computed at a specificity of $95\%$.}
    \label{fig:confusion_matrices}
\end{figure*}
\begin{figure*}[t!]
\centering
    \includegraphics[width=7in,height=2.63in]{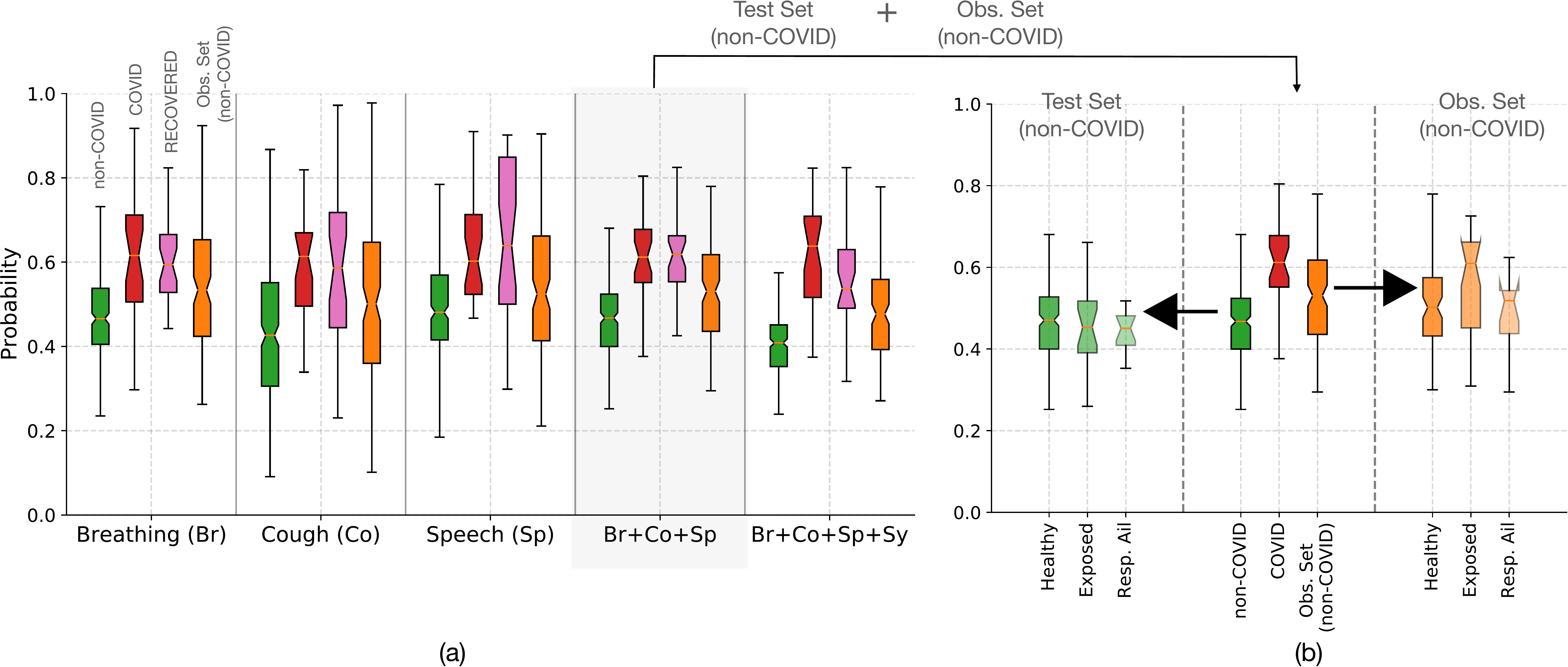}
    \caption{(a) Distribution of the COVID probability score ($p$). Green - non-COVID, Red - COVID Positive, Pink - fully recovered from COVID, Orange - observation set. (b) A refined breakup of the fusion system Br+Co+Sp output with respect to  non-COVID class participants (green and orange).}
    \label{fig:score_distribution}
\end{figure*}

The bottom row in Figure \ref{fig:functionals_summary} depicts the ROCs on the test set using the classifiers trained on the entire development set. The AUC is $0.79$, $0.74$ and $0.79$ for Br, Co and Sp categories, respectively.   Also, the performance of the lin-SVM is similar to LR. Further, the use of non-linear RBF kernel did not show consistent benefits. We use the logistic regression (LR) which is a linear classifier in all the subsequent experiments. 
\subsection{Symptom classifier}
\noindent Figure \ref{fig:estimated_decision_tree} shows the final decision tree model obtained after five-fold cross-validation. At each node, the classifier tests whether a given symptom is present or not. The numerical value at the leaf node is the probability of COVID class. The probability is greater than $0.76$ if any one of the symptoms are present, and is $0.2$ if none of the symptoms are present. For isolated symptoms of cough, cold and fever, the predicted probabilities are $0.775,~0.821$ and $~0.89$ respectively. The isolated symptoms of loss of smell and fatigue are also assigned probability greater than $0.9$ (higher odds ratio seen in Figure ~\ref{fig:symptoms_distribution}). The symptom of sore throat has the smallest probability of $0.764$. The ROC for the decision tree classifier is shown in Figure~\ref{fig:scorefusion_rocs}, where an AUC of $0.80$ is observed. 

\subsection{Multi-modal diagnostics}\label{sec:scorefusionexp}
\begin{table}[t]
    \caption{Cross correlation coefficient of the test scores ($p$) from the classifiers trained on the four categories.}
    \centering
    \begin{tabular}{|c|cccc|}
    \hline
    Category    & Breathing & Cough & Speech & Symptom \\
    \hline
   Breathing & 1.0 & 0.225 & 0.354 & 0.281\\
    Cough & 0.225 & 1.0 & 0.339 & 0.106\\
      Speech & 0.354 & 0.339 & 1.0 & 0.089\\
    Symptom & 0.281 & 0.105 & 0.089 & 1.0\\ \midrule 
    
    \end{tabular}

    \label{tab:score_ncc}
    
    \vspace{-0.25in}
\end{table}

\noindent We investigate the cross-correlation between the test scores predicted by classifiers trained on the four categories of multi-modal data -  three categories of acoustics and the symptom data. Table. \ref{tab:score_ncc} shows the cross-correlation coefficient values on the test data. The correlation coefficient is less than $0.4$ for all the pairs of modalities. The table shows that the scores predicted using symptoms have less correlation with scores from the sound categories of cough and speech.

\noindent We analyze a score fusion across the multi-modal data categories (Equation~\eqref{eq:fusion}). Figure \ref{fig:scorefusion_rocs} shows the test ROCs for the individual modalities, fusion of the acoustic categories, and the fusion of all the four categories. The fusion of the acoustic categories yields an improvement over the individual categories and gives an AUC of $0.84$. The symptom data has better performance over the sound categories in the region of low sensitivity. Further, the fusion of the four categories improves the overall AUC significantly, and gives an AUC of $0.92$. It is noteworthy that this AUC performance is achieved using classical machine learning models, LR and decision-tree, along with an arithmetic mean based score combination. 

\par Figure \ref{fig:functionals_score_combination} shows the test AUC, sensitivity at  specificity of $95$\%, positive predictive value (PPV), and negative predictive value (NPV) for all the individual modalities and the fusion based combination. The figure shows that the combination of acoustic modalities improves the performance on all the measures reported.  The fusion of the three acoustic categories is found to improve the test AUC by $5\%$ points over the best performing individual sound category. The sensitivity values also show improved performance for the fusion of acoustic categories compared to the individual modality.  

\begin{table*}[t!]
\centering
\caption{Test AUC with \text{(fSet)} and without \text{(All\textbackslash fSet)} feature subsets. Feature dimensions are provided in Table \ref{tab:llds}}

\begin{tabular}{|c|cc|cc|cc|}
\hline
Features & \multicolumn{2}{c|}{Breathing} & \multicolumn{2}{c|}{Cough} & \multicolumn{2}{c|}{Speech} \\ \hline
All & \multicolumn{2}{c|}{0.78} & \multicolumn{2}{c|}{0.74} & \multicolumn{2}{c|}{0.79} \\ \hline 
Subset (fSet) Type & fSet & All\textbackslash fSet & fSet & All\textbackslash fSet & fSet & All\textbackslash fSet \\\hline 
RMS Energy, Zero-Crossing Rate & 0.69 & 0.77 & 0.61 & 0.73 & 0.75 & 0.79\\
Sum of modulation filtered auditory spectrum & 0.70 & 0.78 & 0.66 & 0.74 & 0.71 & 0.79\\
Sum of auditory spectrum (loudness) & 0.74 & 0.78 & 0.67 & 0.74 & 0.73 & 0.79\\
All energy features & 0.73 & 0.77 & 0.68 & 0.72 & 0.78 & 0.79\\\hline
modulation filtered auditory spectrogram (0–8 kHz) & 0.73 & 0.74 & 0.55 & 0.70 & 0.73 & 0.78\\
Mel frequency Cepstral coefficients (MFCC) & 0.66 & 0.78 & 0.74 & 0.69 & 0.76 & 0.77\\
Spectral Flux, Centroid, Entropy, Slope & 0.72 & 0.78 & 0.65 & 0.74 & 0.72 & 0.79\\
Psyoacousic Sharpness, Harmonicity & 0.70 & 0.78 & 0.60 & 0.74 & 0.67 & 0.79\\
Spectral Roll-Off Pt. 0.25, 0.5, 0.75, 0.9 & 0.78 & 0.77 & 0.59 & 0.74 & 0.66 & 0.79\\
Spectral Variance, Skewness, Kurtosis & 0.61 & 0.75 & 0.63 & 0.73 & 0.69 & 0.79\\
Spectral energy in bands 250–650 Hz, 1 kHz–4 kHz & 0.62 & 0.78 & 0.60 & 0.75 & 0.62 & 0.80\\
All spectral features & 0.75 & 0.72 & 0.72 & 0.64 & 0.79 & 0.75\\\hline
F0 (SHS \& Viterbi smoothing) & 0.58 & 0.78 & 0.47 & 0.75 & 0.63 & 0.79\\
Log. HNR, Jitter (local, $\Delta$), Shimmer (local) & 0.62 & 0.76 & 0.48 & 0.75 & 0.62 & 0.80\\
Probability of voicing & 0.73 & 0.77 & 0.62 & 0.73 & 0.60 & 0.79\\
All voicing features & 0.68 & 0.75 & 0.50 & 0.75 & 0.70 & 0.80\\\hline
\end{tabular}

\label{tab:feature_importance_analysis}
\end{table*}
Figure \ref{fig:functionals_score_combination} also shows the performance for fusion of symptoms and the acoustic modalities as well as the fusion with pairs of acoustic categories. We see that fusion with symptoms improves the performance of the acoustic based classifiers significantly. The fusion of all the four modalities has the test AUC of $0.92$, an absolute improvement of $8\%$ compared to the fusion of the acoustic categories alone. At $95\%$ specificity, a sensitivity of $69\%$ is achieved for the fusion of all modalities. The corresponding PPV is $0.75$ with a NPV value of $0.95$. 

\par Next, we investigate the decision power of the developed classifiers. Figure \ref{fig:confusion_matrices} shows the confusion matrices for the four modalities and the score fusion. The confusion matrices are shown for an operating point with specificity value of $0.95$.  At the chosen operating point, the overall accuracy   of the fusion system is $92.7\%$, and the class-weighted accuracy   is $82.1\%$. 

\subsection{Score analysis}\label{sec:score_analysis}
\noindent We analyze the score distribution from the individual modalities as well as the fusion system. The score distribution is given in Figure~\ref{fig:score_distribution}(a). The dataset distribution used in training and testing of the models is shown in Figure~\ref{fig:bd_experiment_procedure}.   Figure~\ref{fig:score_distribution}(a) shows the score distribution for the recovered subset (in blue shade) along with the score distribution of COVID class and non-COVID class data (green and red shade respectively). As seen here, for all the classifier settings, the score distribution of the participants in the recovered status is found to be similar to those seen in COVID positive subjects. While the coupling of this data with the duration of the recovery period would have enabled an in-depth analysis, we would like to point out that the majority of the data came from participants who had just been discharged from the hospital facility. This indicates that the acoustic bio-markers of COVID may last for longer periods of time.    

Figure~\ref{fig:score_distribution}(a) also shows the score distribution of the observation set  (set of participants who are non-COVID, but recorded during the second wave period of April 1, 2021 to May 7, 2021. All participants in this observation set came from India). The observation set consists of data from three categories - completely healthy with no exposure to COVID, healthy but exposed to COVID and individuals with pre-existing respiratory ailments but not infected with COVID. The score distribution for this subset is shown in orange shade in Figure~\ref{fig:score_distribution}(a). As seen here, the data from the observation set also tends to have a higher probability score compared to the score values obtained for the non-COVID category. 

A closer analysis is given in Figure~\ref{fig:score_distribution}(b). Here, the score distribution of the blind test data belonging to the non-COVID category is compared with the score distribution of the observation subset (also of non-COVID category). It is observed that the subset of participants who have self-reported as ``exposed'' have a higher value of scores during this period of second wave in India compared to the same category of participants from a previous time period. This indicates that the observation set of participants with ``exposed'' category may have  acoustic bio-markers of the COVID infection.   

\subsection{Acoustic feature ablation experiments}
\noindent The ComParE2016 features~\cite{functionals} are the statistics computed from low-level descriptors of $13$ different categories shown in Table \ref{tab:llds}. In this section, we analyze the importance of subsets of features by (i) training a classifier with a selected feature subset (fSet) and (ii) training a classifier excluding the selected feature subset (\{All features\}\textbackslash\{fSet\}). Table \ref{tab:feature_importance_analysis} shows the test AUC obtained for different subsets. We see that the performance of statistical features based on MFCC alone is comparable to the use of all the features for cough and speech. For cough sounds, excluding MFCC based features from the feature set also results in a degradation in the performance, indicating that statistical features of MFCC capture the essential characteristics for classification of cough sounds. More than one subset of features gives test AUC comparable to the one using the full set for breathing and speech.  However, the subsets with AUC above $0.7$ correspond to the broad category of spectral features.

Since multiple subsets have predictive power for breathing and speech, excluding a specific subset did not degrade the performance.   The score combination of the best-performing feature categories in Table \ref{tab:feature_importance_analysis}, i.e., $400$-dimensional spectral roll-off features for breathing, $1400$-dimensional MFCC features for cough, and the $400$-dimensional energy features for speech, gives a test AUC of $0.88$.  The above analysis suggests that the ComParE2016 features capture redundant information with good predictive power for the COVID classification task.

\section{Discussion}\label{sec:discussion}
\noindent The paper presents a novel approach to point-of-care diagnostics of COVID-19 using multi-modal data of acoustics and symptoms. The data used in this study comes from a web-based crowd-sourced data collection platform. The acoustic features  are based on statistical measures of low-level acoustic descriptors while binary features are extracted from the symptom data. The proposed multi-modal diagnostic tool for COVID (MuDiCoV) is the fusion of scores from individual classifiers.

The classification models are simple linear models like logistic regression and decision tree classifiers having a small memory footprint ($600$~kB). On the test set, the average time from input to decision was found to be $5.06~$secs (std. dev. $1.21$~secs) on a desktop computer with Intel(R) Core(TM) i7-6700 CPU @ 3.40GHz processor and $16$~GB memory. The   performance (with $69$\% sensitivity at $95\%$ specificity) obtained on the test set surpasses the benchmark set by the Indian council of medical research (ICMR) for approval of POCT, ($\ge50\%$ sensitivity at $\ge95\%$ specificity) \cite{icmr_rat}.
We foresee that the use of simple classifiers and models would allow the diagnostic methods to be more interpretable. The proposed methodology combines  all the advantages of being a rapid, low-cost, scalable, and remotely usable testing  approach.




\section*{Acknowledgments}
\noindent The authors would  like to express their gratitude to Anand Mohan for the design of the web based data collection platform. The authors would like to thank Dr. Nirmala, Dr. Shrirama Bhat,  Dr. Chandra Kiran and Dr. Suhail Khalid for their co-ordination in data collection. The authors would  like to acknowledge Amir Poorjam and Flavio Avila for discussions on ComParE2016 features.

\bibliography{refs.bib}
\bibliographystyle{IEEEtran}

\end{document}

%% file: plots/tables/LLDs.tex
\begin{table}[t]
    \centering
      \caption{The set of low-level descriptors computed in the ComParE2016 feature set~\cite{opensmile}.}
    \begin{tabular}{|c|c|c|}
    \hline
     & Low level descriptor (LLD) & Dim. \\\hline
    \multirow{3}{*}{\rotatebox[origin=c]{90}{Energy}}&  RMS Energy, Zero-Crossing Rate & 2\\
    &  Sum of modulation-filtered auditory spectrum & 1\\
    &  Sum of auditory spectrum (loudness) & 1\\\hline
    \multirow{7}{*}{\rotatebox[origin=c]{90}{Spectral}}& modulation-filtered auditory spectrogram (0–8 kHz) & 26\\
    &  Mel frequency Cepstral coefficients (MFCCs) & 14\\
    &  Spectral Flux, Centroid, Entropy, Slope & 4\\
    &  Psychoacoustic Sharpness, Harmonicity & 2 \\
    &  Spectral Roll-Off Pt. 0.25, 0.5, 0.75, 0.9 & 4\\
    &  Spectral Variance, Skewness, Kurtosis & 3 \\
    &  Spectral energy 250–650 Hz, 1 kHz–4 kHz & 2 \\\hline
    \multirow{3}{*}{\rotatebox[origin=c]{90}{Voicing}}&  F0 (SHS \& Viterbi smoothing) & 1\\
    &  Log. HNR, Jitter (local, $\Delta$), Shimmer (local) & 4 \\
    &  Probability of voicing & 1\\\hline
    \end{tabular}
  
    \label{tab:llds}
\end{table}

%% file: plots/tables/Functionals.tex
\begin{table}[t]
    \centering
    \caption{List of statistical features  derived from low-level descriptors given in Table~\ref{tab:llds} \cite{functionals}.}
    \begin{tabular}{|c|c|}
    \hline
{Statistic features derived from LLD}               & {Group}       \\ \hline
Quartiles 1–3, 3 inter-quartile ranges & \multirow{3}{*}{Percentiles} \\ 
$1\%$ percentile ($\sim$ min), $99\%$ percentile ($\sim$ max) &  \\ 
Percentile range 1 \%–99 \%                       & \\ \hline
Position of min / max, range (max – min) & \multirow{7}{*}{Temporal}    \\
Contour centroid, flatness                  &    \\
Rel. duration LLD is $> 25~/~50~/75~/90\%$ range &    \\
Relative duration LLD is rising                   &    \\
Relative duration LLD has positive curvature          &    \\
Mean, max, min, std. deviation of segment length       &    \\
Percentage of non-zero frames                     &    \\ \hline
Mean value of peaks                               & \multirow{6}{*}{Peaks}       \\ 
Mean value of peaks – arithmetic mean             &       \\ 
Mean / std.deviation of inter peak distances           &       \\ 
Amplitude mean of peaks, of minima                &       \\ 
Amplitude range of peaks                          &       \\ 
Mean / std. deviation of rising / falling slopes       &       \\ \hline
Arithmetic mean, root quadratic mean              & \multirow{2}{*}{Moments}     \\ 
Standard deviation, skewness, kurtosis            &  \\ \hline
Linear regression slope, offset, quadratic error  & \multirow{2}{*}{Regression}  \\ 
Quadratic regression a, b, offset, quadratic error &   \\ \hline
Linear prediction (LP) gain, LP Coeffecients 1–5     & {Modulation}  \\ \hline
    \end{tabular}
    
    \label{tab:functionals}
\end{table}

%% file: val_test_plots.tex
\begin{tikzpicture}
\node[inner sep=0pt,anchor=center] (main_fig) at (2,10) 
{\includegraphics[width=2in,height=2in]{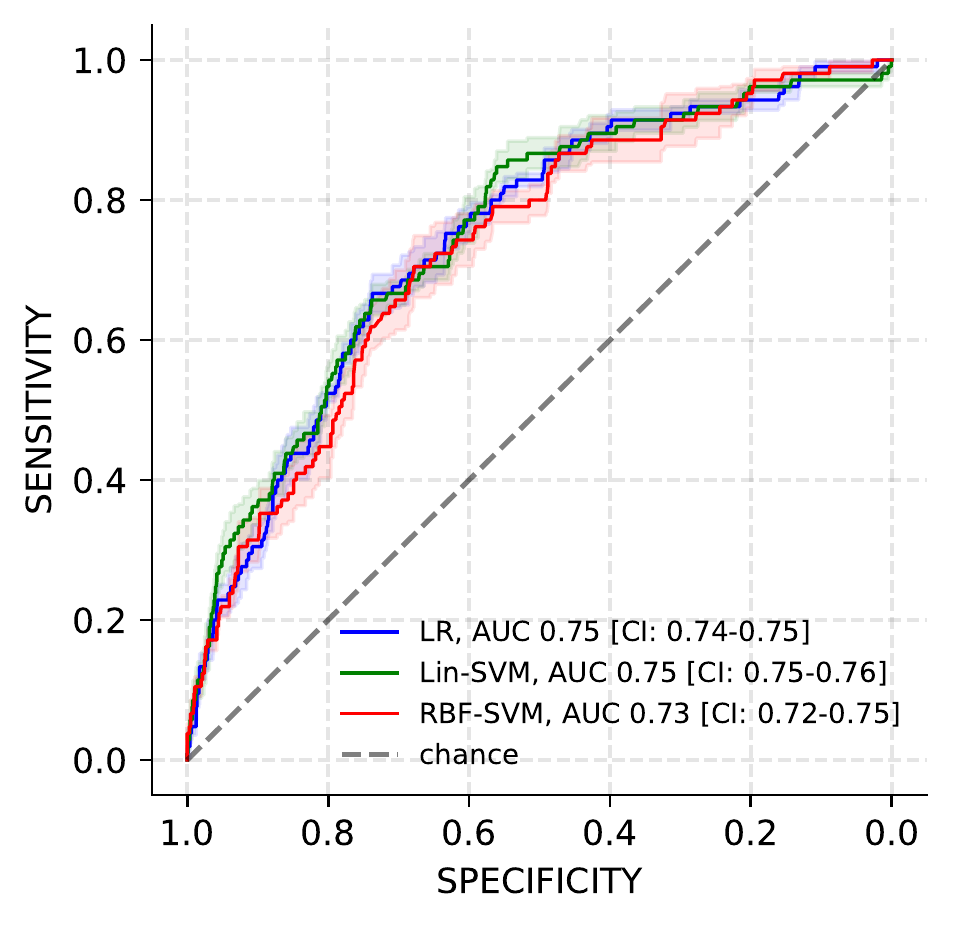}};

\node[inner sep=0pt,anchor=center] (main_fig) at (8,10) 
{\includegraphics[width=2in,height=2in]{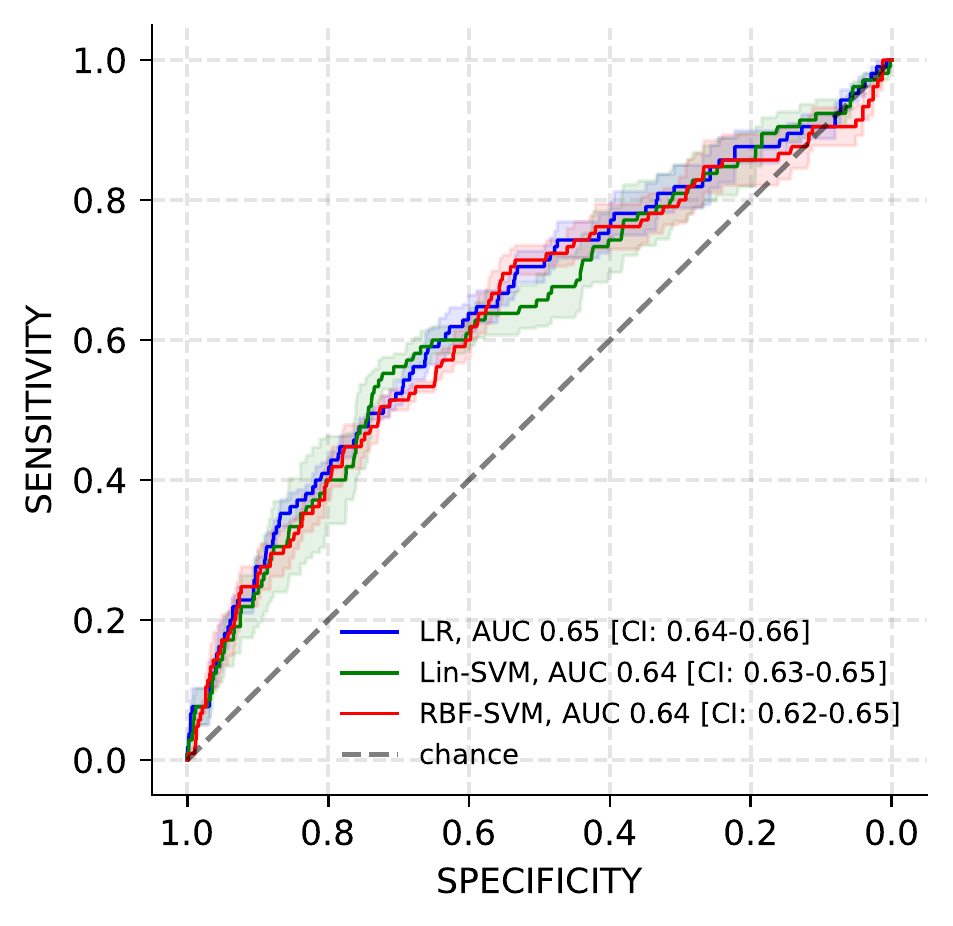}};

\node[inner sep=0pt,anchor=center] (main_fig) at (14,10) 
{\includegraphics[width=2in,height=2in]{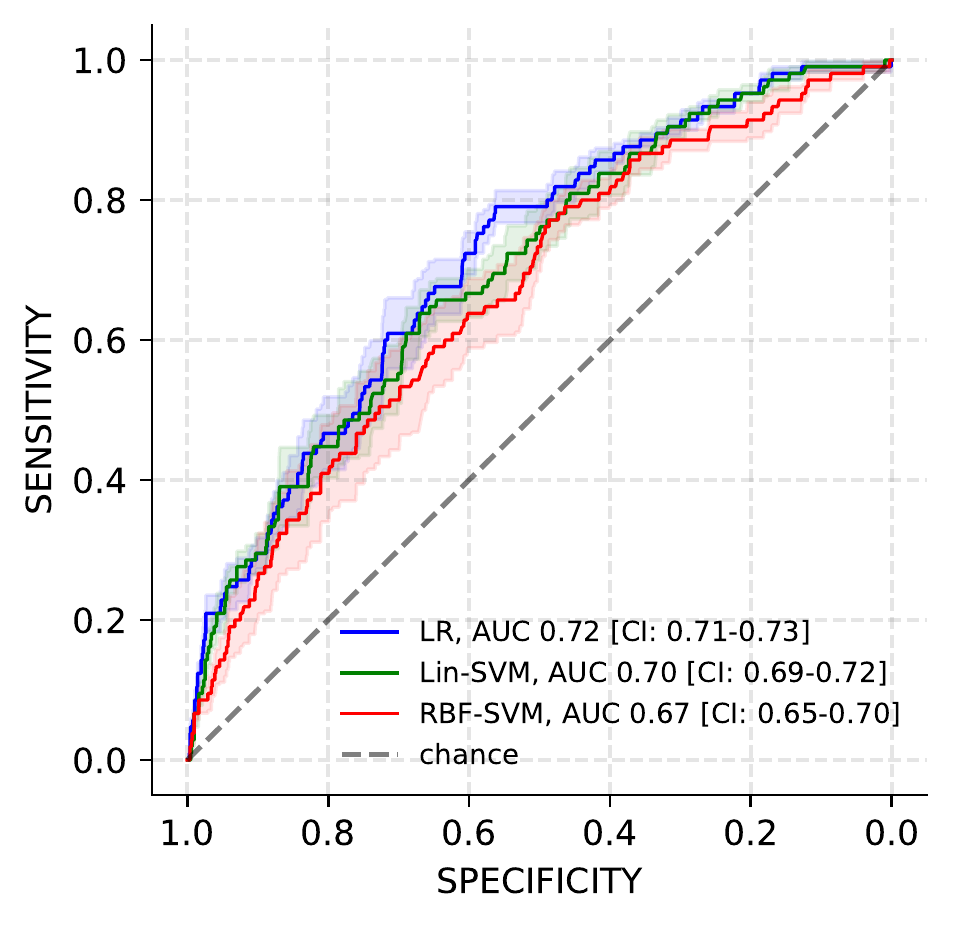}};

\node[inner sep=0pt,anchor=center] (main_fig) at (2,3) 
{\includegraphics[width=2in,height=2in]{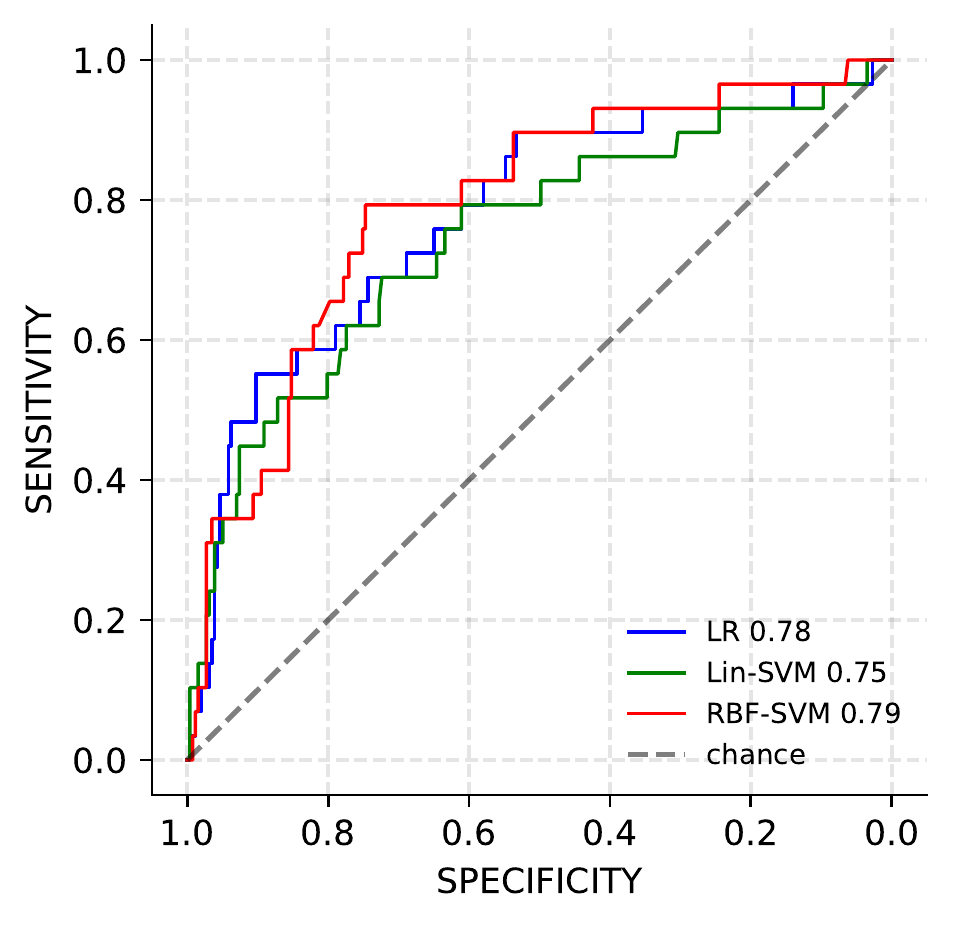}};

\node[inner sep=0pt,anchor=center] (main_fig) at (8,3) 
{\includegraphics[width=2in,height=2in]{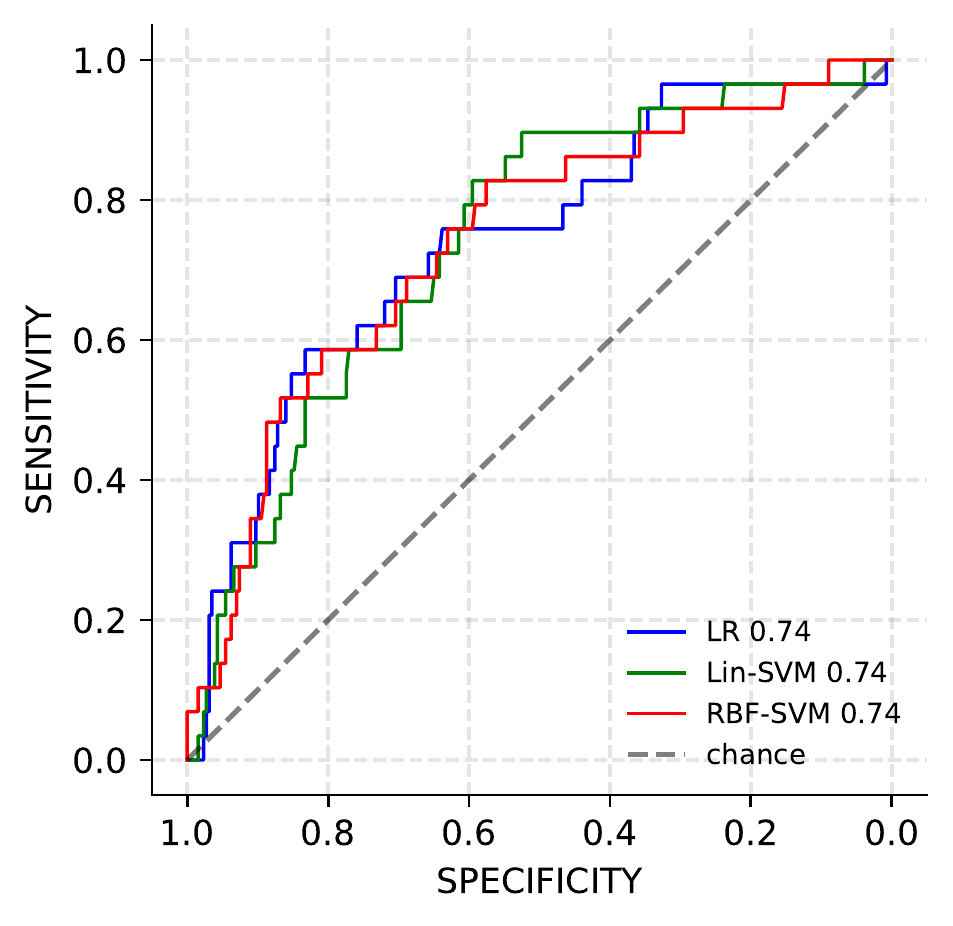}};

\node[inner sep=0pt,anchor=center] (main_fig) at (14,3) 
{\includegraphics[width=2in,height=2in]{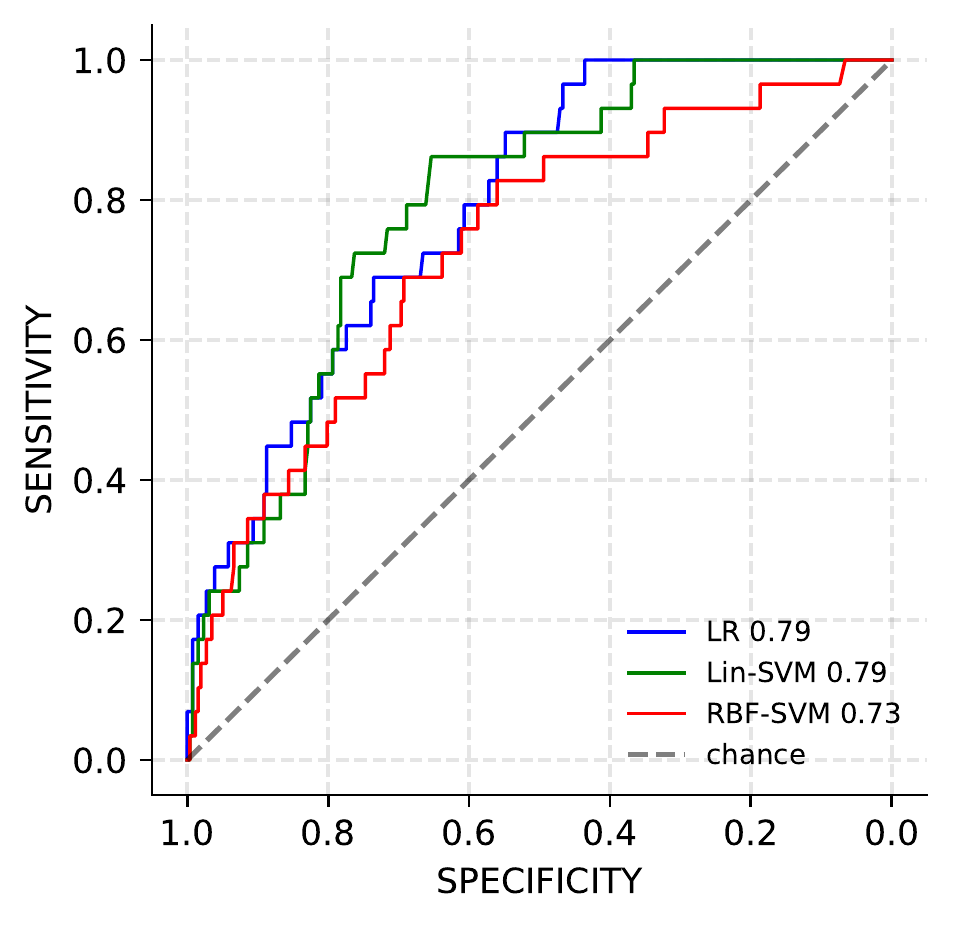}};

\node[font=\fontsize{10}{8}\selectfont,rotate=0,anchor=center] at (8,13.1) (l1) {Performance on Validation folds};
\node[font=\fontsize{8}{6}\selectfont,rotate=0,anchor=center] at (2,12.5) (l1) {Breathing};
\node[font=\fontsize{8}{6}\selectfont,rotate=0,anchor=center] at (8,12.5) (l1) {Cough};
\node[font=\fontsize{8}{6}\selectfont,rotate=0,anchor=center] at (14,12.5) (l1) {Speech};
\node[font=\fontsize{10}{8}\selectfont,rotate=0,anchor=center] at (8,7.2) (l1) {(a)};

\draw[thin, gray, dashed] (0,6.75) -- (16.5,6.75);

\node[font=\fontsize{10}{8}\selectfont,rotate=0,anchor=center] at (8,6.1) (l1) {Performance on Test set};
\node[font=\fontsize{8}{6}\selectfont,rotate=0,anchor=center] at (2,5.5) (l1) {Breathing};
\node[font=\fontsize{8}{6}\selectfont,rotate=0,anchor=center] at (8,5.5) (l1) {Cough};
\node[font=\fontsize{8}{6}\selectfont,rotate=0,anchor=center] at (14,5.5) (l1) {Speech};
\node[font=\fontsize{10}{8}\selectfont,rotate=0,anchor=center] at (8,0.2) (l1) {(b)};

\end{tikzpicture}

    

%% file: plots/decision_tree.tex
\begin{figure}[t]
    \centering
    \resizebox{8.5cm}{4.5cm}{%
\begin{tikzpicture}[scale=0.9, every node/.style={scale=0.9}]
    \node[draw,rounded corners](n1) at (0,0) {Cough};
    \node[draw,rounded corners](n2) at (-1,-1) {Fever};
    \node[draw,rounded corners](n3) at (1,-1) {Fever};  
    \draw[->,black!50!green,thick] (n1) --node[left,black,pos=0.3]{False} (n2);    
    \draw[->,red,thick] (n1) --node[right,black,pos=0.3]{True} (n3);
    \node[draw,rounded corners](n4) at (-3,-2) {Cold};
    \node[draw,rounded corners](n5) at (-1,-2) {Muscle pain};
    \node[draw,rounded corners](n6) at (1,-2) {Fatigue};
    \node[draw,rounded corners](n7) at (3,-2) {Cold};  
    \draw[->,black!50!green,thick] (n2) -- (n4);
    \draw[->,red,thick] (n2) -- (n5);
    \draw[->,black!50!green,thick] (n3) -- (n6);
    \draw[->,red,thick] (n3) -- (n7);

    \node[draw,rounded corners](n8) at (-4.8,-3) {Loss of smell};
    \node[](n9) at (-3,-3) {0.821};
    \node[](n10) at (-1.5,-3) {0.89};
    \node[](n11) at (-0.5,-3) {1.0};  
    \node[draw,rounded corners](n12) at (0.5,-3) {Cold};
    \node[](n13) at (1.5,-3) {1.0};
    \node[draw,rounded corners](n14) at (3.0,-3) {Sore throat};
    \node[](n15) at (4.8,-3) {1.0};  
    
    \draw[->,black!50!green,thick] (n4) -- (n8);
    \draw[->,red,thick] (n4) -- (n9);
    \draw[->,black!50!green,thick] (n5) -- (n10);
    \draw[->,red,thick] (n5) -- (n11);
    \draw[->,black!50!green,thick] (n6) -- (n12);
    \draw[->,red,thick] (n6) -- (n13);
    \draw[->,black!50!green,thick] (n7) -- (n14);
    \draw[->,red,thick] (n7) -- (n15);    
    
    \node[draw,rounded corners](n16) at (-5.7,-4) {Fatigue};
    \node[](n17) at (-4.0,-4) {0.928};    
    \draw[->,black!50!green,thick] (n8) -- (n16);
    \draw[->,red,thick] (n8) -- (n17);
    \node[](n18) at (-0.1,-4) {0.775};
    \node[](n19) at (1.1,-4) {0.883};    
    \draw[->,black!50!green,thick] (n12) -- (n18);
    \draw[->,red,thick] (n12) -- (n19);

    \node[](n20) at (2.4,-4) {0.96};
    \node[](n21) at (3.6,-4) {0.975};    
    \draw[->,black!50!green,thick] (n14) -- (n20);
    \draw[->,red,thick] (n14) -- (n21);    
    
    \node[draw,rounded corners](n22) at (-6.2,-5) {Sore throat};
    \node[](n23) at (-4.5,-5) {0.936};    
    \draw[->,black!50!green,thick] (n16) -- (n22);
    \draw[->,red,thick] (n16) -- (n23);
    
    \node[](n24) at (-6.8,-6) {0.2};
    \node[](n25) at (-5.6,-6) {0.764};    
    \draw[->,black!50!green,thick] (n22) -- (n24);
    \draw[->,red,thick] (n22) -- (n25);    
\end{tikzpicture}
}
    \caption{Decision tree after training on symptoms samples from the development set. The value at the leaf node shows probability score for COVID class.}
    \label{fig:estimated_decision_tree}
    \vspace{-0.2in} 
\end{figure}

%% file: main.bbl
\begin{thebibliography}{10}
\providecommand{\url}[1]{#1}
\csname url@samestyle\endcsname
\providecommand{\newblock}{\relax}
\providecommand{\bibinfo}[2]{#2}
\providecommand{\BIBentrySTDinterwordspacing}{\spaceskip=0pt\relax}
\providecommand{\BIBentryALTinterwordstretchfactor}{4}
\providecommand{\BIBentryALTinterwordspacing}{\spaceskip=\fontdimen2\font plus
\BIBentryALTinterwordstretchfactor\fontdimen3\font minus
  \fontdimen4\font\relax}
\providecommand{\BIBforeignlanguage}[2]{{%
\expandafter\ifx\csname l@#1\endcsname\relax
\typeout{** WARNING: IEEEtran.bst: No hyphenation pattern has been}%
\typeout{** loaded for the language `#1'. Using the pattern for}%
\typeout{** the default language instead.}%
\else
\language=\csname l@#1\endcsname
\fi
#2}}
\providecommand{\BIBdecl}{\relax}
\BIBdecl

\bibitem{hu2020characteristics}
B.~Hu, H.~Guo, P.~Zhou, and Z.-L. Shi, ``Characteristics of {SARS-CoV-2} and
  {{COVID}-19},'' \emph{Nature Reviews Microbiology}, pp. 1--14, 2020.

\bibitem{who_test_test}
``{WHO Director-General's opening remarks at the media briefing on {{COVID}-19
  -16 March 2020}},''
  \url{https://www.who.int/director-general/speeches/detail/who-director-general-s-opening-remarks-at-the-media-briefing-on-{COVID}-19---16-march-2020},
  2020, [Online; accessed 20-May-2021].

\bibitem{corman2020detection}
V.~M. Corman, O.~Landt, M.~Kaiser, R.~Molenkamp, A.~Meijer, D.~K. Chu,
  T.~Bleicker, S.~Br{\"u}nink, J.~Schneider, M.~L. Schmidt \emph{et~al.},
  ``Detection of 2019 novel coronavirus (2019-ncov) by real-time {RT-PCR},''
  \emph{Eurosurveillance}, vol. 25.2000045, no.~3, 2020.

\bibitem{peeling2021scaling}
R.~W. Peeling, P.~L. Olliaro, D.~I. Boeras, and N.~Fongwen, ``Scaling up
  {{COVID}-19} rapid antigen tests: promises and challenges,'' \emph{The Lancet
  infectious diseases}, 2021.

\bibitem{scohy2020low}
A.~Scohy, A.~Anantharajah, M.~Bod{\'e}us, B.~Kabamba-Mukadi, A.~Verroken, and
  H.~Rodriguez-Villalobos, ``Low performance of rapid antigen detection test as
  frontline testing for {{COVID}-19} diagnosis,'' \emph{Journal of Clinical
  Virology}, vol. 129, p. 104455, 2020.

\bibitem{who_poct}
``{Target product profiles for priority diagnostics to support response to the
  {COVID}-19 pandemic v.1.0 (WHO)},''
  \url{https://www.who.int/docs/default-source/blue-print/who-rd-blueprint-diagnostics-tpp-final-v1-0-28-09-jc-ppc-final-cmp92616a80172344e4be0edf315b582021.pdf?sfvrsn=e3747f20_1&download=true},
  2020, [Online; accessed 20-May-2021].

\bibitem{laennec1838treatise}
R.~T.~H. Laennec and J.~Forbes, \emph{A Treatise on the Diseases of the Chest,
  and on Mediate Auscultation}.\hskip 1em plus 0.5em minus 0.4em\relax Samuel
  S. and William Wood, 1838.

\bibitem{pramono2016cough}
R.~X.~A. Pramono, S.~A. Imtiaz, and E.~Rodriguez-Villegas, ``A cough-based
  algorithm for automatic diagnosis of pertussis,'' \emph{PloS one}, vol.~11,
  no.~9, p. e0162128, 2016.

\bibitem{windmon2018tussiswatch}
A.~Windmon, M.~Minakshi, P.~Bharti, S.~Chellappan, M.~Johansson, B.~A. Jenkins,
  and P.~R. Athilingam, ``{TussisWatch:} a smart-phone system to identify cough
  episodes as early symptoms of chronic obstructive pulmonary disease and
  congestive heart failure,'' \emph{IEEE J. Biomedical and Health Informatics},
  vol.~23, no.~4, pp. 1566--1573, 2018.

\bibitem{botha2018detection}
G.~Botha, G.~Theron, R.~Warren, M.~Klopper, K.~Dheda, P.~Van~Helden, and
  T.~Niesler, ``Detection of tuberculosis by automatic cough sound analysis,''
  \emph{Physiological measurement}, vol.~39, no.~4, p. 045005, 2018.

\bibitem{porter2020diagnosing}
P.~Porter, S.~Claxton, J.~Brisbane, N.~Bear, J.~Wood, V.~Peltonen, P.~Della,
  F.~Purdie, C.~Smith, and U.~Abeyratne, ``Diagnosing chronic obstructive
  airway disease on a smartphone using patient-reported symptoms and cough
  analysis: Diagnostic accuracy study,'' \emph{JMIR Formative Research},
  vol.~4, no.~11, p. e24587, 2020.

\bibitem{abeyratne2013cough}
U.~R. Abeyratne, V.~Swarnkar, A.~Setyati, and R.~Triasih, ``Cough sound
  analysis can rapidly diagnose childhood pneumonia,'' \emph{Annals of
  Biomedical engineering}, vol.~41, no.~11, pp. 2448--2462, 2013.

\bibitem{swarnkar2013automatic}
V.~Swarnkar, U.~R. Abeyratne, A.~B. Chang, Y.~A. Amrulloh, A.~Setyati, and
  R.~Triasih, ``Automatic identification of wet and dry cough in pediatric
  patients with respiratory diseases,'' \emph{Annals of biomedical
  engineering}, vol.~41, no.~5, pp. 1016--1028, 2013.

\bibitem{hee2019development}
H.~I. Hee, B.~Balamurali, A.~Karunakaran, D.~Herremans, O.~H. Teoh, K.~P. Lee,
  S.~S. Teng, S.~Lui, and J.~M. Chen, ``Development of machine learning for
  asthmatic and healthy voluntary cough sounds: a proof of concept study,''
  \emph{Applied Sciences}, vol.~9, no.~14, p. 2833, 2019.

\bibitem{li2021epidemiology}
J.~Li, D.~Q. Huang, B.~Zou, H.~Yang, W.~Z. Hui, F.~Rui, N.~T.~S. Yee, C.~Liu,
  S.~N. Nerurkar, J.~C.~Y. Kai \emph{et~al.}, ``Epidemiology of {{COVID}-19}: A
  systematic review and meta-analysis of clinical characteristics, risk
  factors, and outcomes,'' \emph{Journal of Medical Virology}, vol.~93, no.~3,
  pp. 1449--1458, 2021.

\bibitem{HUANG2020497}
C.~Huang, Y.~Wang, X.~Li, L.~Ren, J.~Zhao, Y.~Hu, L.~Zhang, G.~Fan, J.~Xu,
  X.~Gu, Z.~Cheng, T.~Yu, J.~Xia, Y.~Wei, W.~Wu, X.~Xie, W.~Yin, H.~Li, M.~Liu,
  Y.~Xiao, H.~Gao, L.~Guo, J.~Xie, G.~Wang, R.~Jiang, Z.~Gao, Q.~Jin, J.~Wang,
  and B.~Cao, ``Clinical features of patients infected with 2019 novel
  coronavirus in {Wuhan, China},'' \emph{The Lancet}, vol. 395, no. 10223, pp.
  497--506, 2020.

\bibitem{9391791}
K.~Qian, B.~W. Schuller, and Y.~Yamamoto, ``Recent advances in computer
  audition for diagnosing {{COVID}-19}: An overview,'' in \emph{2021 IEEE 3rd
  Global Conf. Life Sciences and Technologies (LifeTech)}, 2021, pp. 181--182.

\bibitem{brown2020exploring}
C.~Brown, J.~Chauhan, A.~Grammenos, J.~Han, A.~Hasthanasombat, D.~Spathis,
  T.~Xia, P.~Cicuta, and C.~Mascolo, ``Exploring automatic diagnosis of
  {COVID}-19 from crowdsourced respiratory sound data,'' in \emph{Proc. 26th
  ACM SIGKDD International Conference on Knowledge Discovery \& Data Mining},
  2020, pp. 3474--3484.

\bibitem{orlandic2020coughvid}
L.~Orlandic, T.~Teijeiro, and D.~Atienza, ``The {COUGHVID} crowdsourcing
  dataset: A corpus for the study of large-scale cough analysis algorithms,''
  \emph{arXiv preprint arXiv:2009.11644}, 2020.

\bibitem{laguarta}
J.~Laguarta, F.~Hueto, and B.~Subirana, ``{{COVID}-19} artificial intelligence
  diagnosis using only cough recordings,'' \emph{IEEE Open Journal of
  Engineering in Medicine and Biology}, vol.~1, pp. 275--281, 2020.

\bibitem{sharma2020coswara}
N.~Sharma, P.~Krishnan, R.~Kumar, S.~Ramoji, S.~R. Chetupalli, R.~Nirmala,
  P.~K. Ghosh, and S.~Ganapathy, ``Coswara -- a database of breathing, cough,
  and voice sounds for {{COVID}-19} diagnosis,'' in \emph{Proc. Interspeech},
  2020, pp. 4811--4815.

\bibitem{coppock2021end}
H.~Coppock, A.~Gaskell, P.~Tzirakis, A.~Baird, L.~Jones, and B.~Schuller,
  ``End-to-end convolutional neural network enables {{COVID}-19} detection from
  breath and cough audio: a pilot study,'' \emph{BMJ Innovations}, vol.~7,
  no.~2, 2021.

\bibitem{agbley2020wavelet}
B.~L.~Y. Agbley, J.~Li, A.~Haq, B.~Cobbinah, D.~Kulevome, P.~A. Agbefu, and
  B.~Eleeza, ``Wavelet-based cough signal decomposition for multimodal
  classification,'' in \emph{17th Intl. Computer Conference on Wavelet Active
  Media Technology and Information Processing)}.\hskip 1em plus 0.5em minus
  0.4em\relax IEEE, 2020, pp. 5--9.

\bibitem{9401826}
K.~Feng, F.~He, J.~Steinmann, and I.~Demirkiran, ``Deep-learning based approach
  to identify {COVID}-19,'' in \emph{SoutheastCon 2021}, 2021, pp. 1--4.

\bibitem{9361107}
J.~Andreu-Perez, H.~Perez-Espinosa, E.~Timonet, M.~Kiani, M.~I. Giron-Perez,
  A.~B. Benitez-Trinidad, D.~Jarchi, A.~Rosales, N.~Gkatzoulis, O.~F.
  Reyes-Galaviz, A.~Torres, C.~Alberto Reyes-Garcia, Z.~Ali, and F.~Rivas, ``A
  generic deep learning based cough analysis system from clinically validated
  samples for point-of-need {{COVID}-19} test and severity levels,'' \emph{IEEE
  Trans. Services Computing}, pp. 1--1, 2021.

\bibitem{imran2020ai4}
A.~Imran, I.~Posokhova, H.~N. Qureshi, U.~Masood, M.~S. Riaz, K.~Ali, C.~N.
  John, M.~I. Hussain, and M.~Nabeel, ``{AI4{COVID}-19: AI} enabled preliminary
  diagnosis for {{COVID}-19} from cough samples via an app,'' \emph{Informatics
  in Medicine Unlocked}, vol.~20, p. 100378, 2020.

\bibitem{9416469}
L.~Verde, G.~De~Pietro, A.~Ghoneim, M.~Alrashoud, K.~N. Al-Mutib, and
  G.~Sannino, ``Exploring the use of artificial intelligence techniques to
  detect the presence of coronavirus {COVID}-19 through speech and voice
  analysis,'' \emph{IEEE Access}, vol.~9, pp. 65\,750--65\,757, 2021.

\bibitem{smell_metadata}
\BIBentryALTinterwordspacing
C.~Menni, A.~M. Valdes, M.~B. Freidin, C.~H. Sudre, L.~H. Nguyen, D.~A. Drew,
  S.~Ganesh, T.~Varsavsky, M.~J. Cardoso, J.~S. El-Sayed~Moustafa, A.~Visconti,
  P.~Hysi, R.~C.~E. Bowyer, M.~Mangino, M.~Falchi, J.~Wolf, S.~Ourselin, A.~T.
  Chan, C.~J. Steves, and T.~D. Spector, ``Real-time tracking of self-reported
  symptoms to predict potential {{COVID}-19},'' \emph{Nature Medicine},
  vol.~26, pp. 1037--1040, 2020. [Online]. Available:
  \url{https://doi.org/10.1038/s41591-020-0916-2}
\BIBentrySTDinterwordspacing

\bibitem{zoabi2021machine}
Y.~Zoabi, S.~Deri-Rozov, and N.~Shomron, ``Machine learning-based prediction of
  {COVID}-19 diagnosis based on symptoms,'' \emph{npj Digital Medicine},
  vol.~4, no.~1, pp. 1--5, 2021.

\bibitem{steepestdescent}
\BIBentryALTinterwordspacing
H.~B. Curry, ``The method of steepest descent for non-linear minimization
  problems,'' \emph{Quarterly of Applied Mathematics}, vol.~2, no.~3, pp.
  258--261, 1944. [Online]. Available:
  \url{http://www.jstor.org/stable/43633461}
\BIBentrySTDinterwordspacing

\bibitem{Platt99probabilisticoutputs}
J.~C. Platt, ``Probabilistic outputs for support vector machines and
  comparisons to regularized likelihood methods,'' in \emph{Advances in Large
  Margin Classifiers}.\hskip 1em plus 0.5em minus 0.4em\relax MIT Press, 1999,
  pp. 61--74.

\bibitem{myles2004introduction}
A.~J. Myles, R.~N. Feudale, Y.~Liu, N.~A. Woody, and S.~D. Brown, ``An
  introduction to decision tree modeling,'' \emph{Journal of Chemometrics: A
  Journal of the Chemometrics Society}, vol.~18, no.~6, pp. 275--285, 2004.

\bibitem{ComParE2016}
B.~Schuller, S.~Steidl, A.~Batliner, J.~Hirschberg, J.~K. Burgoon, A.~Baird,
  A.~Elkins, Y.~Zhang, E.~Coutinho, and K.~Evanini, ``The {INTERSPEECH} 2016
  computational paralinguistics challenge: Deception, sincerity \& native
  language,'' in \emph{Proc. of Interspeech}, 2016, pp. 2001--2005.

\bibitem{opensmile}
F.~Eyben, M.~W\"{o}llmer, and B.~Schuller, ``Opensmile: The munich versatile
  and fast open-source audio feature extractor,'' in \emph{Proc. 18th ACM Intl.
  Conf. Multimedia}, 2010, p. 1459–1462.

\bibitem{functionals}
F.~Weninger, F.~Eyben, B.~Schuller, M.~Mortillaro, and K.~Scherer, ``On the
  acoustics of emotion in audio: What speech, music, and sound have in
  common,'' \emph{Frontiers in Psychology}, vol.~4, p. 292, 2013.

\bibitem{scikit-learn}
F.~Pedregosa, G.~Varoquaux, A.~Gramfort, V.~Michel, B.~Thirion, O.~Grisel,
  M.~Blondel, P.~Prettenhofer, R.~Weiss, V.~Dubourg, J.~Vanderplas, A.~Passos,
  D.~Cournapeau, M.~Brucher, M.~Perrot, and E.~Duchesnay, ``Scikit-learn:
  Machine learning in {P}ython,'' \emph{Journal of Machine Learning Research},
  vol.~12, pp. 2825--2830, 2011.

\bibitem{icmr_rat}
``{ICMR Rapid Antigen Test Kits for COVID-19 (Oropharyngeal / Nasopharyngeal
  swabs) - 28 May 2021},''
  \url{https://www.icmr.gov.in/pdf/covid/kits/List_of_rapid_antigen_kits_28052021.pdf},
  2021, [Online; accessed 29-May-2021].

\end{thebibliography}
